\documentclass[letterpaper,twocolumn,10pt,accepted=2020-11-05]{quantumarticle}
\pdfoutput=1
\usepackage[numbers,sort&compress]{natbib}
\usepackage{amsmath, amssymb, graphicx, bm}
\usepackage[svgnames]{xcolor}
\usepackage{comment}
\usepackage{hyperref}
\DeclareMathOperator{\sech}{sech}

\newcommand{\ket}[1]{| #1 \rangle}
\newcommand{\bra}[1]{\langle #1 |}
\newcommand{\braket}[1]{\langle #1 \rangle}
\newcommand{\hc}{\text{H.c.}}

\newcommand{\eq}[1]{\begin{align}#1\end{align}}
\newcommand{\nn}{\nonumber}
  
\newcommand{\uu}{\mathcal{ U}}

\newcommand{\seq}[1]{\begin{subequations}#1\end{subequations}}

\newcommand{\diag}{\text{diag}}

\usepackage{bm}

\renewcommand{\a}{ a}
\newcommand{\ad}{ a^\dagger}

\DeclareMathOperator{\Ai}{Ai}
\newcommand{\Aip}{\mathrm{Ai'}}

\begin{document}

\title{Fast optimization of parametrized quantum optical circuits}

\author{Filippo M. Miatto}
\affiliation{Institut Polytechnique de Paris}
\affiliation{T\'el\'ecom Paris, LTCI, 19 Place Marguerite Perey 91120 Palaiseau}

\author{Nicol\'as Quesada}
\affiliation{Xanadu, Toronto, ON, M5G 2C8, Canada}
\date{November 5, 2020}
\maketitle

\begin{abstract}
Parametrized quantum optical circuits are a class of quantum circuits in which the carriers of quantum information are photons and the gates are optical transformations. Classically optimizing these circuits is challenging due to the infinite dimensionality of the photon number vector space that is associated to each optical mode.
Truncating the space dimension is unavoidable, and it can lead to incorrect results if the gates populate photon number states beyond the cutoff.
To tackle this issue, we present an algorithm that is orders of magnitude faster than the current state of the art, to recursively compute the exact matrix elements of Gaussian operators and their gradient with respect to a parametrization.
These operators, when augmented with a non-Gaussian transformation such as the Kerr gate, achieve universal quantum computation.
Our approach brings two advantages: first, by computing the matrix elements of Gaussian operators directly, we don't need to construct them by combining several other operators; second, we can use any variant of the gradient descent algorithm by plugging our gradients into an automatic differentiation framework such as TensorFlow or PyTorch.
Our results will find applications in quantum optical hardware research, quantum machine learning, optical data processing, device discovery and device design.
\end{abstract}

\section{Introduction}
In recent years the success of machine learning (ML) in statistics, data science and artificial intelligence has sparked much interest in the fields of quantum information and quantum computing.
Nowadays, quantum machine learning is a term that encapsulates various techniques and ideas --- including both optimization of quantum circuits via ML techniques \cite{schuld2019evaluating,benedetti2019parameterized}, as well as running subroutines of ML algorithms on quantum hardware \cite{dunjko2016quantum, kerenidis2019q, cerezo2020variational, larose2019variational, bravo2019variational}.
As a parallel thread, the quantum optics community has investigated how to compound quantum optical primitives to generate complex quantum states \cite{arrazola2019machine,killoran2019continuous,killoran2019strawberry,steinbrecher2019quantum,quesada2019simulating}. For this program to continue, it becomes necessary to develop new and efficient tools to perform fast simulation of quantum optical circuits. 

Parametrized quantum circuits are information processing devices whose elements depend on one or more parameters, and therefore one can think of them as a particular class of neural networks \cite{romero2017quantum,schuld2020circuit,schuld2019quantum,havlivcek2019supervised,o2016scalable,peruzzo2014variational}.
Among parametrized quantum circuits, quantum optical circuits are those in which the carriers of quantum information are photons and the gates are optical transformations \cite{killoran2019continuous, killoran2019strawberry, steinbrecher2019quantum}.
Working in the optical domain offers several advantages such as high processing speed, no demand for vacuum or cold temperatures and the potential to build compact devices via integrated optics \cite{politi2009integrated,politi2009integrated,zhang2020single,vaidya2019broadband}.

A typical architecture of quantum optical circuits alternates between Gaussian transformations and non-linear ones such as the Kerr gate, which is necessary to achieve universality \cite{lloyd1999quantum}.
Although the evolution of Gaussian states under Gaussian transformations can be computed solely from their finite-dimensional symplectic matrix \cite{serafini2017quantum,weedbrook2012gaussian}, Kerr gates transform Gaussian states into non-Gaussian ones and so they make it necessary to work in Fock space, which is infinite-dimensional and requires one to impose a cutoff.

Optimizing a parametrized quantum optical circuit is a challenging task, because one needs to simulate exactly the evolution of an input state as it propagates through the optical components, as well as the gradient of the output with respect to the circuit parameters.
As each component can be represented in matrix form, the transformation of the initial state is a simple matrix multiplication and as such it is efficient.
However, the difficult part is to generate the transformation matrix from the parameters of the gate: optical transformations (even the fundamental ones) are expressed as matrix exponentials of a linear combination of infinite-dimensional non-commuting operators \cite{barnett2002methods}.
Therefore (unless the exponent is diagonal) one cannot simply truncate the matrix at the exponent and subsequently compute the matrix exponential:
\begin{align}
\mathrm{trunc}(\exp(A)) \neq \exp(\mathrm{trunc}(A)).
\end{align}

For some special transformations such as the displacement \cite{cahill1969ordered} and squeezing operator \cite{kral1990displaced}, exact formulas for their matrix elements are known.
For some other transformations, such as two-mode squeezing and beamsplitters, one can resort to disentangling theorems which allow one to express the exponential in LDU form (i.e., as the product of a lower-triangular, a diagonal and an upper-triangular matrices), which can be safely truncated before multiplication, or alternatively allows the matrix elements to be expressed as polynomials of the parameters of the transformations \cite{ma1990multimode,barnett2002methods,quesada2015very}. Bosonic representations of the special unitary group $SU(n)$ have also been studied in the mathematical physics literature \cite{dhand2015algorithms}: in the quantum optical setting, these are the matrix elements of a passive linear optical transformation corresponding to an interferometer. 
In the context of Franck-Condon transitions, the chemical-physics community \cite{doktorov1977dynamical,gruner1987efficient,berger1998calculation,ezspectrum,huh2011unified,rabidoux2016highly}  has studied the matrix elements of Gaussian transformations where the displacement, squeezing parameters and unitary matrices representing interferometers are real, corresponding to so-called point transformations where canonical positions and momenta of the harmonic oscillator do not mix \cite{moshinsky1971linear,wolf1974canonical,kramer1975complex}.  Until this work, no exact method was known for more general Gaussian transformations and typically one expressed them as a product of fundamental truncated gates, accepting that truncation errors could propagate.

Our method to perform the optimization of a parametrized quantum optical circuit has several advantages over previous techniques:
\begin{enumerate}
	\item It generates recursively the matrix elements of general Gaussian transformations exactly up to the desired cutoff dimension without the need for decompositions. These results are derived in Sec.~\ref{sec:gen} for general Gaussian gates and applied to specific cases in Sec.~\ref{sec:apps}.
	\item It constructs gradients exactly, without the need for computational graphs describing the functional dependencies between the variables. This is explained in Sec.~\ref{sec:grads}.
	\item It is numerically stable and orders of magnitude faster than previous methods. This is detailed in Sec.~\ref{sec:exp} where we report on several numerical experiments.
\end{enumerate}

Even though we focus on Gaussian gates in the main text, in Appendices \ref{sec:comms}, \ref{app:cubic}, \ref{app:quartic} and \ref{app:single} we derive the analytical form of the matrix elements of the cubic- and quartic- phase gates together with recurrence relations for their calculation and the calculation of their gradients. These operators are non-Gaussian gates that, unlike the Kerr gate, are not diagonal in the Fock basis.\\
The algorithms we present in this manuscript are part of the \verb|fock_gradients|
module of \texttt{The Walrus} \cite{gupt2019walrus} since version 0.12 released in Dec. 2019. \\
Finally, note that in this work we construct recursively the matrix representation of a Gaussian operator. If instead one were interested in just a few specific matrix elements, one could map this to the calculation of loop hafnians \cite{bjorklund2018faster,quesada2019franck,banchi2020training}.

\section{The generating function method} \label{sec:gen}
\subsection{Notation and definitions}
We summarize here our notational conventions, which are meant to simplify the presence of multiple indices in the multimode formulas, and the definition of our generating function.

We adopt the following short-hand notation:
\eq{
	\bm{\beta}^{\bm{n}} &\equiv  \prod_{i=1}^\ell \beta_i^{n_i}, \label{eq:beta}\\
	\bm{n}! &\equiv \prod_{i=1}^\ell n_i!.\label{eq:n!}
}
We indicate tensor products of $\ell$-mode coherent and Fock states as follows:
\eq{
	\ket{\bm{\beta}} &= \ket{\beta_1}\otimes\ldots\otimes\ket{\beta_\ell},\\
	\ket{\bm{n}} &= \ket{n_1}\otimes\ldots\otimes\ket{n_\ell},
}
which we can use for instance to express a multimode coherent state:
\eq{\label{eq:coherent_multimode}
	\ket{\bm\beta} &= e^{-\frac12||\bm\beta||^2} \sum_{\bm n = \bm 0}^{\bm\infty} \frac{\bm\beta^{\bm n}}{\sqrt{\bm n!}}  \ket{\bm n},
}
{where we used the shorthand notation $\sum_{\bm n = \bm 0}^{\bm\infty}  = \sum_{n_1=0}^\infty \ldots \sum_{n_\ell=0}^\infty$. }
Note that tensor indices are numbered starting from 1 (as in Eq.~\eqref{eq:beta}) while the indices themselves have values that begin from 0 (as in Eq.~\eqref{eq:coherent_multimode}).
We also use interchangeably the following two notations for derivatives, depending on convenience and on the available space:
\begin{align}
\frac{\partial^{\bm k}}{\partial{\bm\beta}^{\bm k}} = \partial_{\bm \beta}^{\bm k} \equiv \prod_i\frac{\partial^{k_i}}{\partial{\beta_i}^{k_i}}.
\end{align}

The key insight of the generating function method is that when we compute the inner product $\langle\bm{\alpha}^*|\mathcal{G}|\bm{\beta}\rangle$ of any operator 
$\mathcal{G}$ (not just a Gaussian one)  with a pair of coherent states, we obtain a function of the coherent amplitudes that can be treated as a generating function for the matrix elements of $\mathcal{G}$:
\begin{align}
\label{Gammadef}
\Gamma(\bm{\alpha},\bm{\beta}) &= e^{\frac12(||\bm{\alpha}||^2+||\bm{\beta}||^2)}\langle \bm{\alpha}^*|\mathcal{G}|\bm{\beta} \rangle \\
&= \sum_{\bm{m},\bm{n}=\bm{0}}^{\bm{\infty}} \frac{\bm{\alpha}^{\bm{m}} \bm{\beta}^{\bm{n}}}{\sqrt{\bm{m}!\bm{n}!}}
\langle \bm{m} |\mathcal{G}|\bm{n} \rangle.
\end{align}
The matrix element $\braket{\bm{m}|\mathcal{G}|\bm{n}}$ is therefore contained in the coefficient of the factor $\bm{\alpha}^{\bm{m}} \bm{\beta}^{\bm{n}}$ in the Taylor series and we can isolate it by computing the derivatives of the appropriate order, at zero:
\begin{align}
\mathcal{G}_{\bm{m} \bm{n}} \equiv \braket{\bm{m}|\mathcal{ G}|\bm{n}}= \frac{\partial^{\bm{m}}_{\bm{\alpha}}\partial^{\bm{n}}_{\bm{\beta}} \Gamma(\bm{\alpha},\bm{\beta}) }{\sqrt{\bm m!\bm n!}}\bigg|_{\bm{\alpha}=\bm{\beta}=0}.
\end{align}
{We will derive recursion relations for $\mathcal{G}_{\bm{mn}}$ directly, instead of calculating the numerator and denominator separately in the ratio in the last equation. Thus our methods are numerically more stable than previous ones where divisions by factorials are common.}

We now {write the amplitudes and Fock indices as direct sums of the bra and ket labels }
\eq{\label{directsum}
	\bm{\nu} &= \left[ \begin{array}{c}
		\bm{\alpha} \\
		\bm{\beta}
	\end{array} \right] \in \mathbb{C}^{2 \ell},\\
	\bm{k} &= \left[ \begin{array}{c}
		\bm{m} \\
		\bm{n}
	\end{array} \right] \in \mathbb{N}_0^{2 \ell},
}
so that we can express all of our calculations in terms of a single vector of amplitudes, including the generating function, which we now indicate as 
\eq{
	\Gamma(\bm \nu) = \sum_{\bm{k}=\bm{0}}^{\bm{\infty}} \frac{\bm{\nu}^{\bm{k}}}{\sqrt{\bm{k}!}} \mathcal{G}_{\bm{k}}.
}
where we take advantage of the definition of the vector $\bm{k}$ in Eq.~\eqref{directsum} and write  $\mathcal{G}_{\bm{k}}$ for  $\mathcal{G}_{\bm{m} \bm{n}}$.
As shown in Appendix \ref{app:gaussian}
, the generating function $\Gamma(\bm\nu)$ for a Gaussian operator $\mathcal G$ is in exponential form $\Gamma(\bm\nu) = C\,e^{Q(\bm\nu)}$. This implies that we can express high-order derivatives recursively, using only the derivatives at zero of the exponent $Q(\bm\nu)$, as shown in subsection \ref{subsec:matrix_elements}. Furthermore, the exponent $Q(\bm\nu)$ is a polynomial of degree 2, which we can write as:
\eq{
	Q(\bm\nu) = \bm\mu^T\bm\nu - \frac{1}{2}\bm\nu^T\bm\Sigma\bm\nu.
}
This means that all of its mixed derivatives of order higher than 2 vanish:
\begin{align}
\partial^{\bm k}_{\bm\nu} Q(\bm\nu)|_{\bm\nu=\bm 0} = 0  \text{ if } \sum_{i=1}^{2\ell} k_i > 2.
\end{align}
Because of this, all the derivatives of the exponent that we will ever need are those of degree 1 and 2:
\eq{
	\frac{\partial}{\partial \nu_i} Q(\bm\nu)|_{\bm{\nu}=\bm{0}}&=\mu_i,\\
	\frac{\partial^2}{\partial \nu_i\partial \nu_j} Q(\bm\nu)|_{\bm{\nu}=\bm{0}}&=-\Sigma_{ij}.
}

Finally, note that although $\mathcal{G}_{\bm k}$ is a tensor of rank $2\ell$, for familiarity with the terminology we still refer to its elements as ``matrix elements''.

\subsection{Generating function for multimode Gaussian transformations}
As a corollary of the Bloch-Messiah decomposition \cite{bloch1962canonical,serafini2017quantum}, a general $\ell-$mode Gaussian transformation can be parametrized as
\begin{align}\label{eq:bm}
\mathcal{G}=\mathcal{G}(\bm{\gamma}, \bm{W}, \bm{\zeta}, \bm{V})	 = \mathcal{ D }(\bm{\gamma}) \mathcal{ U}(\bm{W}) \mathcal{ S }(\bm{\zeta}) \mathcal{ U}(\bm{V}),
\end{align}
where we used the notation $\mathcal{ D }(\bm{\gamma}) = \bigotimes_{i=1}^\ell  D_i(\gamma_i)$, $\mathcal{ S }(\bm{\zeta}) = \bigotimes_{i=1}^\ell  S_i(\zeta_i)$ for tensor products of single-mode displacement and squeezing operators defined by
\eq{
	D_j(\gamma_j) &=  \exp\left(\gamma_j \ad_j-\hc \right), \quad  \gamma_j \in \mathbb{C}, \\
	S_j(\zeta_j)&=\exp\left( \tfrac{\zeta_j^*}{2}  a_j^{2} -\hc\right), \quad  \zeta_j = r_j e^{i \delta_j} \in \mathbb{C} ,
}
{where $r_j$ and $\delta_j$ are the phase and modulus of the complex squeezing parameter $\zeta_j$. }

{In the last equation we used} the annihilation $a_j$ and creation $a_j^\dagger$ operators of the $\ell$ modes satisfying the canonical commutation relations
\eq{
	[a_j,a_l^\dagger] = \delta_{j,l}, \quad [a_j,a_l]= [\ad_j,\ad_l] = 0.
}
The Hilbert space operator $\mathcal{ U}(\bm{V})$ represents a general multimode passive transformation (physically corresponding to an interferometer)
and is parametrized by a unitary matrix $\bm{V}$ that transforms the creation operators as
\eq{\label{mapping}
	\uu^\dagger(\bm{V}) \ad_i \uu(\bm{V})=\sum_{l=1}^\ell V^*_{i l}\a_l^\dagger, \quad 
	\uu(\bm{V}) \ad_l \uu^\dagger(\bm{V})=\sum_{i=1}^\ell V_{i l}\a_i^\dagger.
}

Using these definitions, in Appendix \ref{app:gaussian} we show the following:
\eq{\label{eq:exp}
	&e^{\frac12\left[||\bm{\alpha}||^2+||\bm{\beta}||^2 \right] }\braket{\bm{\alpha}^* | \mathcal{ D }(\bm{\gamma}) \mathcal{ U}(\bm{W}) \mathcal{ S }(\bm{\zeta}) \mathcal{ U}(\bm{V})| \bm{\beta}} = \nn	\\
	& \quad 
	C \exp\left( \bm{\mu}^T \bm{\nu} - \frac12 \bm{\nu}^T \bm{\Sigma} \bm{\nu} \right),	
}
where
\eq{
	&C = \frac{ \exp\left(-\tfrac{1}{2} \left[  ||\bm{\gamma}||^2  + \bm{\gamma}^\dagger  \bm{W} \diag(e^{i \bm{\delta}}  \tanh \bm{r}) \bm{W}^T  \bm{\gamma}^*\right]\right)}{\sqrt{\prod_{i=1}^\ell \cosh r_i}},\\
	&\bm{\mu}^T =\\
	& \left[\bm{\gamma}^\dagger \bm{W} \diag(e^{i \bm{\delta}}  \tanh \bm{r}) \bm{W}^T+\bm{\gamma}^T, -\bm{\gamma}^\dagger \bm{W} \diag(\sech  \bm{r}) \bm{V} \right],		\nn \\
	&\bm{\Sigma} = \\
	&\left[	\begin{array}{c|c}
		\bm{W} \diag(e^{i \bm{\delta}} \tanh \bm{r}) \bm{W}^T &  	-\bm{W} \diag(\sech  \bm{r}) \bm{V}\\
		\hline
		-\bm{V}^T \diag(\sech  \bm{r}) \bm{W^T} & - \bm{V}^T \diag(e^{-i \bm{\delta}}\tanh \bm{r}) \bm{V}
	\end{array}
	\right]. \nonumber
}
Note that 
$\bm \mu$ is a vector of dimension $2\ell$. The matrix $\Sigma$, on the other hand is built out of four $\ell\times\ell$ blocks and has therefore dimension $2\ell\times2\ell$.

With these results we are now ready to write explicit recurrence relations for the matrix elements of a general multimode Gaussian gate.

\subsection{From generating functions to matrix elements}
\label{subsec:matrix_elements}
We use the recent result in  \cite{miatto2019recursive} to express the matrix elements of the Gaussian operator recursively (see Appendix \ref{app:rec}):
\eq{
	\mathcal G_{\bm 0} &=C,\\
	\mathcal G_{\bm k + 1_i} &= \frac{1}{\sqrt{k_i+1}}\sum_{\bm j = \bm 0}^{\bm k} \sqrt{\frac{\bm k!}{\bm j!}}\frac{1}{(\bm k - \bm j)!}\mathcal G_{\bm j}\partial_{\bm \nu}^{\bm k + 1_i - \bm j}Q(\bm\nu)|_{\bm\nu=\bm 0},\label{eq:recrel}
}
where $1_i$ is a vector with all zeros and a single 1 at position $i$, and therefore to obtain the vector $\bm k + 1_i$ we take the vector $\bm k$ and we increment $k_i$ by 1.

We stress that Eq.~\eqref{eq:recrel} works for \emph{any} generating function in exponential form. However, as mentioned above, in our case the exponent $Q$ is a polynomial of degree 2, so the recurrence relation simplifies considerably because the derivative of $Q$ is zero unless $\bm j = \bm k$ or $\bm j = \bm k - 1_l$ for some index $l$:
\eq{
	\mathcal{G}_{\bm k + 1_i} = \frac{1}{\sqrt{k_i+1}}\left(\mathcal{G}_{\bm k}\mu_i - \sum_{l = 1}^{2\ell}\sqrt{k_l}\mathcal{G}_{\bm k - 1_l}\Sigma_{il}\right).\label{eq:recfinal}
}
This, together with Eq.~\eqref{eq:gradfinal} which shows how to obtain gradients, is the primary result of our work. As an example of Eq.~\eqref{eq:recfinal}, we derive the two recurrence relations for the single mode case, i.e., $\ell=1$.
For convenience and ease of vectorization when programming, we can use the first relation to build the first column of the matrix and the second relation to subsequently build the rows:
\eq{
	\mathcal{G}_{m+1,n} =& \frac{1}{\sqrt{m+1}}\biggl(\mathcal{G}_{m,n}\mu_1 - \sqrt{m}\mathcal{G}_{m - 1, n}\Sigma_{11}\nonumber\\
	&- \sqrt{n}\mathcal{G}_{m, n-1}\Sigma_{12}\biggr),\label{eq:1mode recrel 1}\\
	\mathcal{G}_{m,n+1} =& \frac{1}{\sqrt{n+1}}\biggl(\mathcal{G}_{m,n}\mu_2 - \sqrt{m}\mathcal{G}_{m - 1, n}\Sigma_{21}\nonumber\\
	&- \sqrt{n}\mathcal{G}_{m, n-1}\Sigma_{22}\biggr).\label{eq:1mode recrel 2}
}
We represent this example pictorially in Fig.~\ref{fig:Gmatrix_1mode}.

For Gaussian operators on $\ell$ modes, the recurrence relations allow us to combine $2\ell+1$ neighbouring elements in $2\ell$ different ways (by incrementing each of the $2\ell$ indices) and generate $2\ell$ new elements.

Note that these recurrence relations can be rewritten in an alternative form by using the mode operators:
\begin{align}
a_i\mathcal{G} &= \mu_i\mathcal{G} - \left(\sum_{l=1}^{\ell}a_l^\dagger\Sigma_{il}\right)\mathcal{G}- \mathcal{G}\left(\sum_{l=1}^{\ell}a_{l}\Sigma_{i,l+\ell}\right),\\
\mathcal{G}a_{i}^\dagger &= \mu_{i+\ell}\mathcal{G} - \left(\sum_{l=1}^{\ell}a_l^\dagger\Sigma_{i+\ell,l}\right)\mathcal{G}- \mathcal{G}\left(\sum_{l=1}^{\ell}a_{l}\Sigma_{i+\ell,l+\ell}\right).
\end{align}
With these compact relations we can formulate the commutation relations between a general multimode Gaussian operator and the mode operators $a_i$ and $a_i^\dagger$, for instance for a single-mode Gaussian operator, we have:
\begin{align}
[\mathcal{G}, a] &= - \mu_1\mathcal{G} + (\Sigma_{12}+1)\mathcal{G}a  + \Sigma_{11} a^\dagger \mathcal{G},\\
[\mathcal{G}, a^\dagger] &= \mu_2\mathcal{G} - (\Sigma_{21}+1)a^\dagger\mathcal{G}  - \Sigma_{22} \mathcal{G}a.
\end{align}

As an example, we can write the special cases of the commutation relations between the mode operators and the squeezing and displacement operators (using the values from subsections \ref{sec:singlemode_squeezing} and \ref{sec:singlemode_displacement}):
\begin{align}
[S(\zeta), a] &=(1-\mathrm{sech}(r))S(\zeta)a+e^{i\delta}\tanh(r)a^\dagger S(\zeta), \\
[S(\zeta), a^\dagger] &= (\mathrm{sech}(r)-1)a^\dagger S(\zeta) + e^{-i\delta}\tanh(r)S(\zeta)a,\\
[D(\gamma), a] &= -\gamma D(\gamma),\\
[D(\gamma), a^\dagger] &= -\gamma^* D(\gamma).
\end{align}

\section{Applications: single- and two-mode gates and passive Gaussian transformations}\label{sec:apps}
All Gaussian single- and two-mode transformations will be expressed as a combination of four elementary operations: single-mode phase shifts, single-mode squeezing, single-mode displacement and beamsplitters.
These elements are parametrized as follows: the phase rotation gate $R(\phi)$ depends on an angle $\phi$, the single-mode squeezer $S(\zeta)$ depends on a single complex parameter which we express in polar form $\zeta=r e^{i\delta}$, the displacement gate $D(\gamma)$ depends on a single complex parameter $\gamma$ which we leave as is, and finally the beamsplitter $B(\theta, \varphi)$ depends on two angles $(\theta, \varphi)$.
Each gate is expressed as the exponential of mode operators. In particular, for the rotation and beamsplitter gates we write
\begin{align}
R(\phi) &= \exp\left(i\phi\, a^\dagger a\right),\\
B(\theta, \varphi) &= \exp\left[\theta\left(e^{i\varphi}a_1 a_2^\dagger - e^{-i\varphi}a_1^\dagger a_2\right)  \right].
\end{align}
For example, the 50/50 beam splitter with real coefficients corresponds to $(\theta, \varphi)=(\pi/4, 0)$. Note that the last two gates are special cases of a passive linear optical transformation, thus they can be written as $\mathcal{U}(\bm{W})$ where
\seq{
	\eq{
		\bm{W} &= e^{i \phi} \label{phase},	\\
		\bm{W} &= \begin{bmatrix}
			\cos \theta & - e^{-i \varphi} \sin \theta \\
			e^{i \varphi} \sin \theta & \cos \theta
		\end{bmatrix},
	}
}
respectively. 
\subsection{Single-mode Gaussian gate}

\begin{figure}[h!]
	\centering
	\includegraphics[width=0.6\columnwidth]{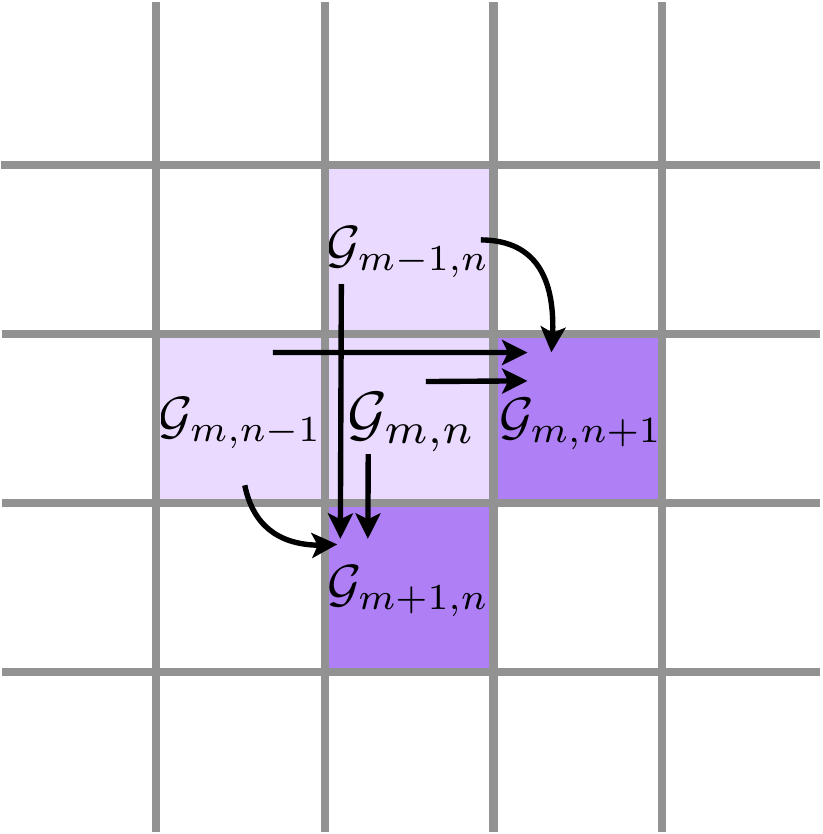}
	\caption{The three neighbouring matrix elements shown in light purple ($\mathcal{G}_{m-1,n}$, $\mathcal{G}_{m,n}$ and $\mathcal{G}_{m,n-1}$) can be linearly combined in two different ways to generate $\mathcal{G}_{m+1,n}$ or $\mathcal{G}_{m,n+1}$. These recurrence rules are also applicable at the edge of the matrix by considering ``outer elements'' as zeros. For Gaussian operators on $\ell$ modes, the recurrence rules allow us to combine $2\ell+1$ neighbouring elements in $2\ell$ different ways and generate $2\ell$ new elements.}
	\label{fig:Gmatrix_1mode}
\end{figure}

A simple way to parametrize a general single-mode Gaussian operation is as follows:
\eq{
	\mathcal{G}^{(1)}= \mathcal{G}^{(1)}(\gamma,\phi,\zeta) = D(\gamma)R(\phi)S(\zeta).
}
By using Eq.~\eqref{phase} and the general result from the last section we easily find
\eq{
	C &=  \frac{ \exp\left(-\tfrac{1}{2} \left[  |\gamma|^2  + \gamma^{*2} e^{i (\delta+ 2\phi)}  \tanh r\right]\right)}{ \sqrt{\cosh r} },\\
	\bm{\mu}^T &= [\gamma^* e^{i (\delta+2 \phi)} \tanh r  + \gamma, -\gamma^* e^{i \phi} \sech r], \\
	\bm{\Sigma} &= \left[ \begin{array}{c|c}
		e^{i(\delta + 2\phi)} \tanh r & - e^{ i \phi} \sech  r \\ 
		\hline
		- e^{ i \phi} \sech  r &  -e^{-i \delta} \tanh r\\
	\end{array}	\right].
}
These values enter the recurrence relations \eqref{eq:1mode recrel 1} and \eqref{eq:1mode recrel 2}, which are depicted in Fig.~\ref{fig:Gmatrix_1mode}.

\subsubsection{Special case: single-mode squeezing}
\label{sec:singlemode_squeezing}
For a single-mode squeezing operator, we have $\phi=0$ and $\gamma=0$, and thus
\eq{
	C &= \sqrt{\sech r },\\
	\bm\mu^T &= [0,0],\\
	\bm\Sigma &= \left[ \begin{array}{c|c}
		e^{i \delta} \tanh r & - \sech  r \\ 
		\hline
		- \sech  r &  -e^{-i \delta} \tanh r\\
	\end{array}	\right].
}
The recurrence relations simplify accordingly, and we obtain the matrix elements of the single-mode squeezer $S_{m ,n} = \langle m |S(\zeta)|n \rangle$:
\begin{align}
S_{0,0} =& \sqrt{\sech r },\\
S_{m+1,0} =& -\sqrt{\frac{m}{m+1}}S_{m-1,0} e^{i\delta}\tanh r,\\
S_{m,n+1} =& \frac{1}{\sqrt{n+1}}\bigl(\sqrt{m}S_{m-1,n} \sech r +\nn\\ &\sqrt{n}S_{m,n-1}e^{-i\delta} \tanh r\bigr),
\end{align}
where we have set $n=0$ in the first recurrence relation so that it builds the first column, and we can use the second relation to build the rows.

\begin{figure}[h!]
	\centering
	\includegraphics[width=0.6\columnwidth]{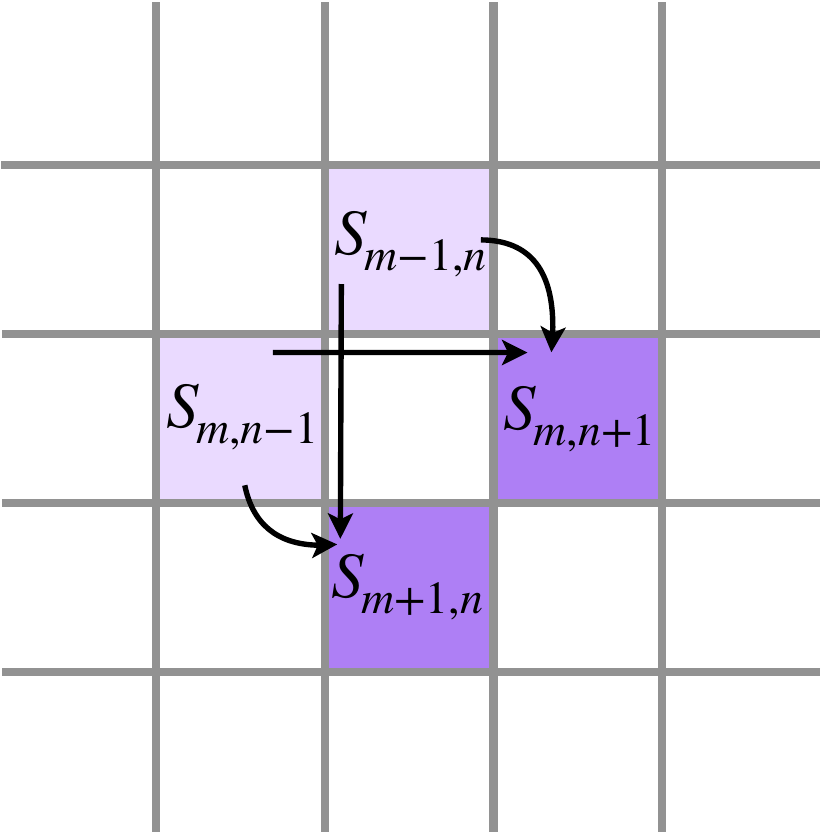}
	\caption{When generating the single-mode squeezer matrix, two matrix elements are sufficient to generate another two at each step. Notice how this produces the checkerboard pattern of zeros that is typical of the squeezer matrix.}
	\label{fig:Smatrix_1mode}
\end{figure}

Thanks to $\bm\mu$ being zero, we do not need to include the middle matrix element at each step while filling out the matrix, as depicted in Fig.~\ref{fig:Smatrix_1mode}. This gives rise to the typical checkerboard pattern of the squeezer matrix.

\subsubsection{Special case: displacement}
\label{sec:singlemode_displacement}
\begin{figure}[h!]
	\centering
	\includegraphics[width=0.6\columnwidth]{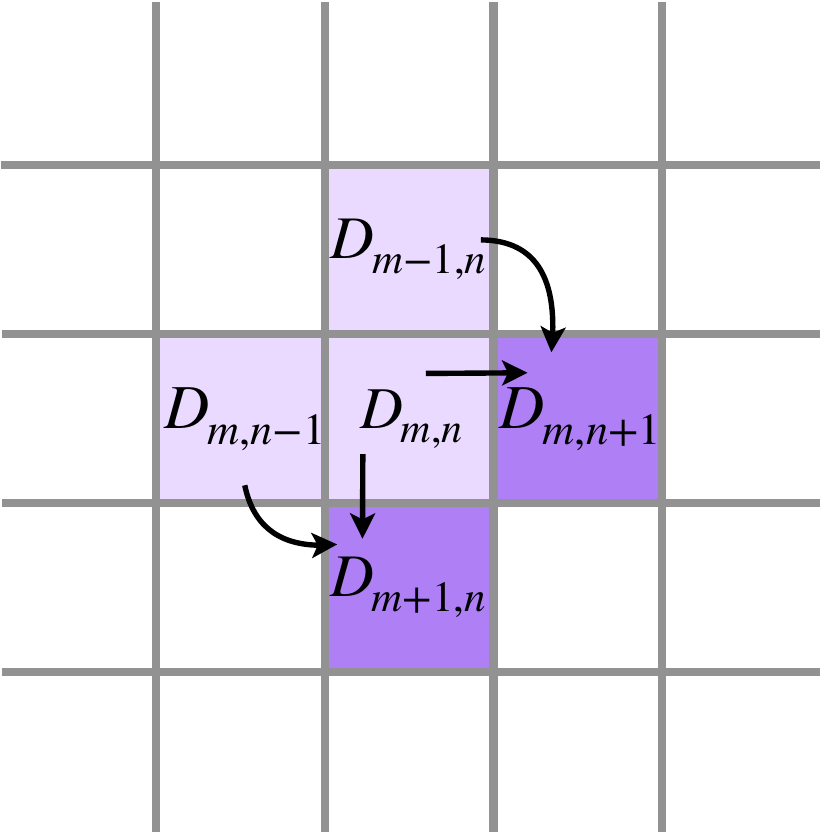}
	\caption{When generating the single-mode displacement matrix, we only need two matrix elements to generate each new element along columns or rows.}
	\label{fig:Dmatrix_1mode}
\end{figure}

For a single-mode displacement operator, we have $\phi=0$ and $\zeta=0$ and therefore $\tanh r=0$ and $\sech r = 1$ and we find
\eq{
	C &= e^{-\frac{1}{2}|\gamma|^2},\\
	\bm\mu^T &= [\gamma,-\gamma^*],\\
	\bm\Sigma &= \left[ \begin{array}{c|c}
		0 & - 1 \\ 
		\hline
		- 1 &  0\\
	\end{array}	\right].
}
The recurrence relations simplify accordingly, and we obtain the matrix elements of the single-mode displacement operator $D_{m,n} = \langle m|D(\gamma)|n \rangle$:
\begin{align}
D_{0,0} &= e^{-\frac{|\gamma|^2}{2}},\\
D_{m+1,0} &= \frac{\gamma}{\sqrt{m+1}}D_{m,0},\\
D_{m,n+1} &= -\frac{\gamma^*}{\sqrt{n+1}}D_{m,n} + \sqrt{\frac{m}{n+1}} D_{m-1,n}.
\end{align}
Thanks to the diagonal of $\bm\Sigma$ being zero, also in this case each step is simplified, as depicted in Fig.~\ref{fig:Dmatrix_1mode}.

\subsection{Two-mode Gaussian gate}
We can write the most general two-mode Gaussian transformation as 
\begin{align}
\mathcal{G}^{(2)}&=\mathcal{D}(\bm \gamma) \mathcal{R}(\bm \phi) B(\theta',\varphi')\mathcal{S}(\bm \zeta)B(\theta,\varphi),
\end{align}
where $\bm{\phi}=[\phi_1,\phi_2]$, $\bm{\gamma}=[\gamma_1,\gamma_2]$ and $\bm{\zeta}=[\zeta_1,\zeta_2]$ and where the single-mode gates factorize over the two modes.
This gives a total of 14 real parameters (counting double for complex parameters), in accordance with Ref.~\cite{cariolaro2016bloch}.
The equations above can be recast into the form of Eq.~\eqref{eq:bm} by writing $\mathcal{U}(\bm{V}) = B(\theta, \varphi)$ and $\mathcal{U}(\bm{W}) =  \mathcal{R}(\bm{\phi}) B(\theta', \varphi')$, with
\eq{\label{2x2unitary}
	\bm{V} &= 	\left[
	\begin{array}{cc}
		\cos  \theta  & -e^{-i \varphi } \sin \theta  \\
		e^{i \varphi } \sin \theta  & \cos \theta  \\
	\end{array}
	\right],\\
	\bm{W} &= \left[
	\begin{array}{cc}
		e^{i \phi _1} & 0 \\
		0 & e^{i \phi _2} \\
	\end{array}
	\right] \left[
	\begin{array}{cc}
		\cos  \theta'  & -e^{-i \varphi' } \sin \theta'  \\
		e^{i \varphi' } \sin \theta'  & \cos \theta'  \\
	\end{array}
	\right].
}


\subsubsection{Special case: Two-mode squeezer}
The two-mode squeezer is defined as $S^{(2)}(\zeta) = \exp\left(\zeta \ad_1\ad_2 - \hc \right)$ and can be decomposed in terms of beamsplitters and single-mode squeezing operations as follows:
\eq{
	S^{(2)}(\zeta) = B(-\pi/4,0) \left [S(\zeta) \otimes S(-\zeta) \right] B(\pi/4,0).
}
From the decomposition above one easily finds $C = \sech r$, $\bm{\mu}^T = [0,0,0,0] $ and
\eq{
	&\bm{\Sigma} = (-1) \times \\ 
	&\left[
	\begin{array}{cccc}
		0 & e^{i \delta } \tanh r & \sech r & 0 \\
		e^{i \delta } \tanh r & 0 & 0 & \sech r \\
		\sech r & 0 & 0 & -e^{-i \delta } \tanh r \\
		0 & \sech r & -e^{-i \delta } \tanh r & 0 \\
	\end{array}
	\right]. \nn
}
Note that a two-mode squeezing operation creates photons in pairs, which implies that the difference of photon number between the two modes is conserved. This observation is mathematically equivalent to the fact that that the operation has an $SU(1,1)$ symmetry \cite{klimov2009group}. If we write $S^{(2)}_{m,n,p,q} = \braket{m,n|S_2(\zeta)|p,q}$ then the only nonzero elements of this tensor satisfy $m-n = p-q$. Matrix elements that do not satisfy this selection rule are zero and we can construct the whole rank-4 tensor by looping over at most 3 indices:

\eq{
	S^{(2)}_{0,0,0,0} =& \sech r,\\
	S^{(2)}_{n,n,0,0} =& - S^{(2)}_{n-1,n-1,0,0}\Sigma_{2,1},\\
	S^{(2)}_{m,n,m-n,0} =& - \sqrt{\frac{m}{m-n}} S^{(2)}_{m-1,n,m-n-1,0}\Sigma_{3,1},\\
	S^{(2)}_{m,n,p,p-(m-n)} =&  \frac{-1}{\sqrt{p-(m-n)}} \times \\
	&\ \biggl( \sqrt{n}S^{(2)}_{m,n-1,p,p-(m-n)-1}\Sigma_{4,2}\nn\\
	&+ \sqrt{p} S^{(2)}_{m,n,p-1,p-(m-n)-1}\Sigma_{4,3}\biggr). \nn
}

\subsection{Passive Gaussian transformations: Interferometers}
These $\ell$-mode transformations, which correspond to interferometers parametrized by an $\ell \times \ell$ unitary matrix $\bm{V}$, can be obtained from our general results by setting $\bm{\zeta} =\bm{\delta}  =  \bm{0}$ and selecting, without loss of generality, $\bm{W} = \mathbb{I}$ to obtain 
\eq{
	C&= 1,\\
	\bm{\mu}^T &= \bm{0},\\
	\bm{\Sigma} &= - \left[\begin{array}{c|c}
		\bm{0} & \bm{V} \\
		\hline
		\bm{V}^T &  \bm{0}
	\end{array}\right].
}
We can write $\braket{\bm{m}| \mathcal{ U} (\bm{V})| \bm{\bm{n}}} = U_{\bm{m},\bm{n}}$ and find the following recurrence relations
\seq{
	\eq{
		{U}_{\bm{0},\bm{0}} &= 1,\\
		{U}_{\bm{m}+1_i,\bm{n}} &= \frac{1}{\sqrt{m_i + 1}} \sum_{j=1}^{\ell} V_{i,j} \sqrt{n_j} \ {U}_{\bm{m},\bm{n} -1_j }, \\
		{U}_{\bm{m},\bm{n}+1_i} &= \frac{1}{\sqrt{n_i + 1}} \sum_{j=1}^{\ell} V_{j,i} \sqrt{m_j} \ {U}_{\bm{m}-1_j,\bm{n}  }.
	}
}
These equations have a particular simple physical interpretation: they tell us that to scatter off one additional photon in the bra side $\bm{m}+1_i$ one photon has to be sent into the interferometer in the ket side $\bm{n}-1_j$. In particular, we find that the probability amplitude that a single photon injected in port $j$ ends up in port $i$ is precisely the $(i,j)$ entry of the unitary matrix describing the interferometer:
\eq{
	{U}_{1_i, 1_j} 	 = V_{i,j}.
}
Note that ${U}_{\bm{m},\bm{n}}$ is nonzero only if $\sum_{i=1}^\ell m_i = \sum_{i=1}^\ell n_i$ since interferometers do not create or destroy particles (mathematically, this is because they have $U(n)$ symmetry).
Matrix elements that do not satisfy this selection rule are zero and we can construct the whole rank-$2\ell$ tensor by looping over at most $2\ell-1$ indices.  Let us illustrate this for a beamsplitter, which is a special case of a general passive transformation, and for which $\bm{V}$ is of size $2\times 2$ and  given in Eq.~\eqref{2x2unitary}. The recurrence relations we find are:
\eq{
	B_{0,0,0,0} =& 1,\\
	B_{m,n,m+n,0} =& \frac{1}{\sqrt{m+n}} \biggl(
	\sqrt{m} V_{1,1}  B_{m-1,n,m+n-1,0} \nn \\
	& \quad \quad + \sqrt{n} V_{2,1} B_{m,n-1,m+n-1,0} \biggr),\\
	B_{m,n,p,m+n-p} =& \frac{1}{\sqrt{m+n-p}} \times   \\
	& \biggl( \sqrt{m} V_{1,2}  B_{m-1,n,p,m+n-p-1} \nn \\
	&+ \sqrt{n} V_{2,2} B_{m,n-1,p,m+n-p-1} \biggr), \nn
}
where $\braket{m,n|B(\theta,\phi)|p,q} = B_{m,n,p,q}$.

Whenever a gate has a selection rule, not only its computation is sped up, but also its application when updating a tensor representing a state. For example, if we write the state of two modes up to cutoff $N$ as
\eq{
	\ket{\psi} = \sum_{k,l=0}^{N-1} c_{k,l} \ket{k,l},
}
then the state after the application of a beamsplitter is given by
\eq{
	\ket{\psi'} =& B(\theta,\phi) \ket{\psi} = \sum_{m,n=0}^{N-1} c'_{n,m} \ket{n,m},
}
and because of the $U(2)$ symmetry of the beamsplitter one can obtain each entry of the updated tensor $c'$ using a single sum:
\eq{
	c'_{n, m} = \sum\limits_{k = \text{max}(1+n+m-N, 0)}
	^{\text{min}(n+m, N-1) + 1}
	B_{n,m,k,n+m-k}  \ c_{k, n+m-k}.
}
Similarly, one can use the $SU(1,1)$ symmetry of the two-mode squeezer to show that if $\ket{\psi''} = \sum_{nm=0}^{N-1} c''_{n,m}\ket{n,m} = S^{(2)}(z)\ket{\psi}$ then the coefficients can again be updated in a single sum as follows:
\eq{
	c''_{n, m} = \sum\limits_{k = \text{max}(n-m, 0)}^{\text{min}(n-m, 0) + N}
	S^{(2)}_{n,m,k, k+m-n} \ c_{k, k+m-n}.
}
These computational savings are doubled when considering density matrices, where updating a single element of a two-mode state  requires in general four sums, but these sums can be cut in half for the beamsplitter or the two-mode squeezer.

\section{Recursive gradients from generating functions}\label{sec:grads}

In this section we present recurrence relations to construct the gradients of Gaussian operators with respect to their parameters. Gradients are essential in optimization procedures, such as the optimization of a whole circuit.

In order to carry out the optimization, we define a function that evaluates the cost or ``loss'' associated with the current parameter values. By design, the optimal choice of parameters corresponds to the global minimum of the loss function. Our goal is then to minimize the loss by slowly varying the parameters in a sequence of steps via a gradient descent algorithm.
In order to apply the gradient descent  algorithm, we need to backpropagate the gradient of the loss function using the chain rule until we reach each of the parameters. Doing so gives us the rate of change of the loss with respect to the parameters, which we can use to update their values and decrease the loss. The optimization procedure starts by initializing the parameters randomly (usually with very small values) and then slowly varies them until we obtain a circuit that is similar enough to the optimal one.

In the next subsections we explain how to correctly handle the gradient of complex and real parameters, we show that gradients can be easily computed from the matrix representation of the operators and we give some examples. 

\subsection{Gradients with respect to complex parameters}
For a complex parameter $\xi$, the gradient descent update step should use the partial derivative of a real loss function $L$ with respect to the \emph{conjugate} of the parameter \cite{hunger2007introduction, guo2020scheme}:
\begin{align}
\xi \leftarrow \xi - \gamma\frac{\partial L}{\partial \xi^*}  .  
\end{align}
Clearly, this update rule falls back to the regular rule in the case of a real parameter.
To compute the gradient for the update, we need to treat complex variables and their conjugate as \emph{independent} variables, which allows us to compute gradients of non-holomorphic functions \cite{hunger2007introduction}. The chain rule then looks as follows:
\begin{align}
\label{chainComplex}
\frac{\partial L}{\partial \xi^*} &= \sum_{\bm k} \frac{\partial L}{\partial \mathcal{G}_{\bm k}^*}\frac{\partial \mathcal{G}_{\bm k}^*}{\partial \xi^*} + \frac{\partial L}{\partial \mathcal{G}_{\bm k}}\frac{\partial \mathcal{G}_{\bm k}}{\partial \xi^*}.
\end{align}
In an automatic differentiation framework such as TensorFlow or PyTorch, if we wish to customize the computation of the gradient of a new operation (e.g., when the new operation makes use of compiled code which is treated like a black box) we are supplied with the upstream gradient tensor $\partial L/\partial \mathcal{G}_{\bm k}^*$. It is our task to combine it with the local gradients $\partial \mathcal{G}_{\bm k}/\partial \xi^*$ and $\partial \mathcal{G}_{\bm k}^*/\partial \xi^*$ as prescribed by the chain rule, to produce the downstream gradient $\partial L/\partial\xi^*$.

In order to obtain the various parts of Eq.~\eqref{chainComplex} we proceed as follows: for the first part of \eqref{chainComplex}, the upstream gradient tensor $\partial L/\partial \mathcal{G}_{\bm k}^*$ is given to us by the software so we only have to compute the local gradient:
\eq{
	\frac{\partial\mathcal{G}_{\bm k}^*}{\partial \xi^*} = \left(\frac{\partial\mathcal{G}_{\bm k}}{\partial \xi}\right)^*.
}
For the second part of \eqref{chainComplex} we can conjugate the upstream gradient tensor (as the loss function is real):
\begin{align}
\frac{\partial L}{\partial \mathcal{G}_{\bm k}} = \left(\frac{\partial L}{\partial \mathcal{G}_{\bm k}^*}\right)^*,
\end{align}
but as the matrix elements of $\mathcal{G}_{\bm k}$ are complex and non-holomorphic functions of $\xi$, their derivatives with respect to $\xi$ and $\xi^*$ are independent and we need to compute $\partial\mathcal{G}_{\bm k}/\partial\xi^*$ separately.
On the other hand, for a real parameter $x$ computing a single gradient tensor $\frac{\partial \mathcal{G}_{\bm k}}{\partial x}$ would suffice:
\begin{align}
\frac{\partial L}{\partial x} &= \sum_{\bm k} \frac{\partial L}{\partial \mathcal{G}_{\bm k}^*}\frac{\partial \mathcal{G}_{\bm k}^*}{\partial x} + \frac{\partial L}{\partial \mathcal{G}_{\bm k}}\frac{\partial \mathcal{G}_{\bm k}}{\partial x}\\
&=2\Re\left(\sum_{\bm k} \frac{\partial L}{\partial \mathcal{G}_{\bm k}}\frac{\partial \mathcal{G}_{\bm k}}{\partial x}\right),
\end{align}
as in this case $\frac{\partial \mathcal{G}_{\bm k}^*}{\partial x}$ and $\frac{\partial \mathcal{G}_{\bm k}}{\partial x}$ are the conjugate of each other.
If one prefers to have only real parameters then there is an additional step: one can write each complex parameter in Cartesian or polar form, and then treat each of these as an additional function of two real variables, e.g., for $\xi=re^{i\phi}$:
\eq{
	\frac{\partial L}{\partial r} &=2\Re\left(\frac{\partial L}{\partial \xi^*}\frac{\partial \xi^*}{\partial r}\right)=2\Re\left(\frac{\partial L}{\partial \xi^*}e^{-i\phi}\right),\\
	\frac{\partial L}{\partial \phi} &=2\Re\left(\frac{\partial L}{\partial \xi^*}\frac{\partial \xi^*}{\partial \phi}\right)=-2\Re\left(\frac{\partial L}{\partial \xi^*}i\xi^*\right).
}
Furthermore, if the gate depends on a single parameter (real or complex) more simplifications can be made as shown in Appendix \ref{app:single}.
\begin{figure*}[ht!]
	\centering
	\includegraphics[width=0.8\textwidth]{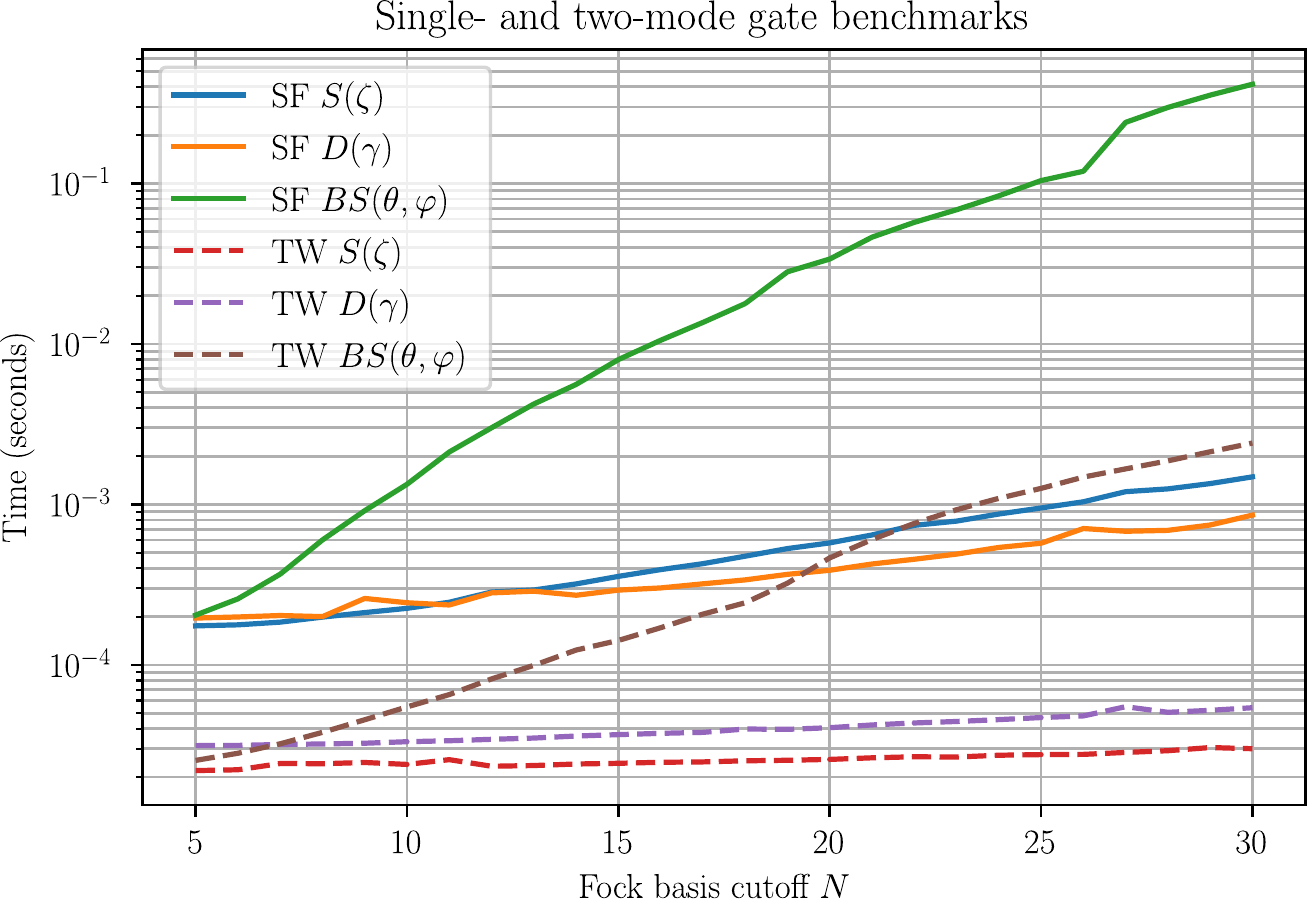}
	\caption{Benchmarks for the computation of the squeezing $S(\zeta)$, displacement $D(\gamma)$ and beamsplitter $B(\theta,\varphi)$ gates using \texttt{Strawberry Fields} (SF)  0.12.1 and the methods from this work as implemented in \texttt{The Walrus} (TW). Note that for a cutoff of $N=30$ our implementations are two orders of magnitude faster. Finally, we also benchmarked the two-mode squeezing gate; the times we find for our implementation are visually indistinguishable from the ones for our beamsplitter, reflecting the fact that both gates can be implemented by looping over three indices despite being rank-4 tensors.
		The benchmarks in this figure were performed on a single core of an Intel-i5 processor @ 2.50 GHz.
	}
	\label{fig:sfvstw}
\end{figure*}
\subsection{Computing gradients with the generating function method}
An $\ell$-mode Gaussian gate has $2\ell^2 + 3\ell$ real parameters (counting double for complex parameters), therefore in order to obtain all the parameter updates we need to compute $2\ell^2 + 3\ell$ derivatives, some of which will be combined to form the parameter updates of the complex parameters.

Another key insight of the generating function method is that to compute the gradients of the matrix elements we can simply differentiate the generating function with respect to the parameters \emph{before} computing the high-order derivatives (here with respect to a generic parameter $\xi$):
\begin{align}
\partial_\xi \mathcal{G}_{\bm k} &= \frac{\partial_{\bm \nu}^{\bm k}}{\sqrt{\bm k!}}\partial_\xi\Gamma(\bm \nu)|_{\bm\nu = \bm 0} = \frac{\partial_{\bm \nu}^{\bm k}}{\sqrt{\bm k!}}\partial_\xi C e^{Q(\bm\nu)}|_{\bm\nu = \bm 0}\\
&= \frac{\partial_{\bm \nu}^{\bm k}}{\sqrt{\bm k!}}\Gamma(\bm \nu) \left(\frac{\partial_\xi C}{C} + \partial_\xi Q(\bm \nu)\right)|_{\bm\nu=\bm 0}\\
&=\left.  \frac{\partial_{\bm \nu}^{\bm k}}{\sqrt{\bm k!}}\Gamma(\bm \nu) \left(\frac{\partial_\xi C}{C} + \frac{\partial\bm\mu^T}{\partial\xi}{\bm \nu}-\frac12{\bm\nu}^T\frac{\partial\bm\Sigma}{\partial\xi}{\bm\nu}\right)\right|_{\bm\nu=\bm 0}\label{eq:gradients_quadratic} .
\end{align}
Expression \eqref{eq:gradients_quadratic} contains terms proportional to $\bm \nu^{\bm n}\Gamma(\bm \nu)$, whose derivative can be expressed using the elements of $\mathcal G_{\bm k}$:
\begin{align}
\frac{\partial^{\bm k}_{\bm \nu}}{\sqrt{\bm k!}} {\bm \nu}^{\bm n}\Gamma(\bm \nu)|_{\bm \nu=\bm 0} = \sqrt{(\bm k)_{\bm n}}\mathcal{G}_{\bm k-\bm n},
\end{align}
where $(k)_{n} = k(k-1)(k-2)\dots(k-n+1)$ is the Pochhammer symbol and $(\bm k)_{\bm n} = \prod_{i}(k_i)_{n_i}$ (with the convention that $(k)_0 = 1$).
So once we compute the tensor representation of a Gaussian operator, we can use it to compute its gradient with respect to a parametrization.
Continuing from Eq.~\eqref{eq:gradients_quadratic} we can write:
\eq{
	\frac{\partial \mathcal{G}_{\bm k}}{\partial\xi} =& \frac{\partial_\xi C}{C}\mathcal{G}_{\bm k} + \sum_i \frac{\partial\mu_i}{\partial\xi}\sqrt{k_i}\mathcal{G}_{\bm k - 1_i}\nonumber \\
	&- \sum_{i>j} \frac{\partial\Sigma_{ij}}{\partial\xi}\sqrt{k_ik_j}\mathcal{G}_{\bm k - 1_i - 1_j}\nonumber\\
	&- \frac{1}{2}\sum_{i} \frac{\partial\Sigma_{ii}}{\partial\xi}\sqrt{k_i(k_i-1)}\mathcal{G}_{\bm k - 2_i}.\label{eq:gradfinal}
}
This, together with Eq.~\eqref{eq:recfinal}, is the primary result of our work.

As an example, we show the derivatives of the single mode Gaussian operator with respect to the complex displacement parameter $\gamma$ and $\gamma^*$ (recall that we must treat them as independent variables). We begin by differentiating $C$, $\bm \mu$ and $\bm \Sigma$:
\eq{
	\frac{\partial_\gamma C}{C} & = -\frac{\gamma^*}{2},\\
	\frac{\partial_{\gamma^*} C}{C} & = -\frac{\gamma}{2} - \gamma^*e^{i(\delta+2\phi)}\tanh r,\\
	\frac{\partial\bm \mu}{\partial \gamma} &= [1,0], \\
	\frac{\partial\bm \mu}{\partial \gamma^*} &= [e^{i(\delta+2\phi)}\tanh r, -e^{i\phi}\sech r],\\
	\frac{\partial\bm \Sigma}{\partial \gamma} &= \bm 0,\\
	\frac{\partial\bm \Sigma}{\partial \gamma^*} &= \bm 0.
}

Therefore, the gradients of the matrix elements are:
\begin{align}
\frac{\partial\mathcal{G}_{m,n}}{\partial \gamma} =& - \frac{\gamma^*}{2}\,\mathcal{G}_{m,n} + \sqrt{m}\,\mathcal{G}_{m-1,n} , \\
\frac{\partial\mathcal{G}_{m,n}}{\partial \gamma^*} =& - \left(\frac{\gamma}{2} +\frac{1}{2}\gamma^*e^{i(\delta+2\phi)}\tanh r\right)\mathcal{G}_{m,n} \\
&
+ e^{i(\delta+2\phi)}\sqrt{m}\mathcal{G}_{m-1,n} \tanh r \nn \\
&- e^{i\phi}\sqrt{n}\mathcal{G}_{m,n-1} \sech r. \nn
\end{align}
We then combine both of these expressions with the upstream gradient $\partial L/\partial \mathcal{G}_{mn}^*$ and its conjugate as prescribed in Eq.~\eqref{chainComplex} to obtain the single gradient for the update, $\partial L/\partial \gamma^*$. We proceed in an analogous way for all the other parameters.

\section{Numerical experiments}\label{sec:exp}
In this section we present a series of experiments that show the effectiveness of our method. First, we benchmark our methods and implementations for the displacement, squeezing and beamsplitter gate against the implementations of \texttt{Strawberry Fields} (SF) version 0.12.1. We note that as of version 0.13, SF uses the methods presented here and implemented in \texttt{The Walrus}. The results of the benchmarks are summarized in Fig.~\ref{fig:sfvstw}; our implementations are two orders of magnitude faster than the ones in SF 0.12.1, and are also more stable and memory efficient. For example, if one tries to calculate a displacement gate by using its known expansion in terms of Laguerre polynomials \cite{cahill1969ordered} it is necessary to perform divisions by the factorial of the number of photons $n$, which results in numerical overflows for $n\sim 100$. Our explicit recurrence relations avoid completely this issue for any number of modes.
\begin{figure}[t!]
	\includegraphics[width=0.95\columnwidth]{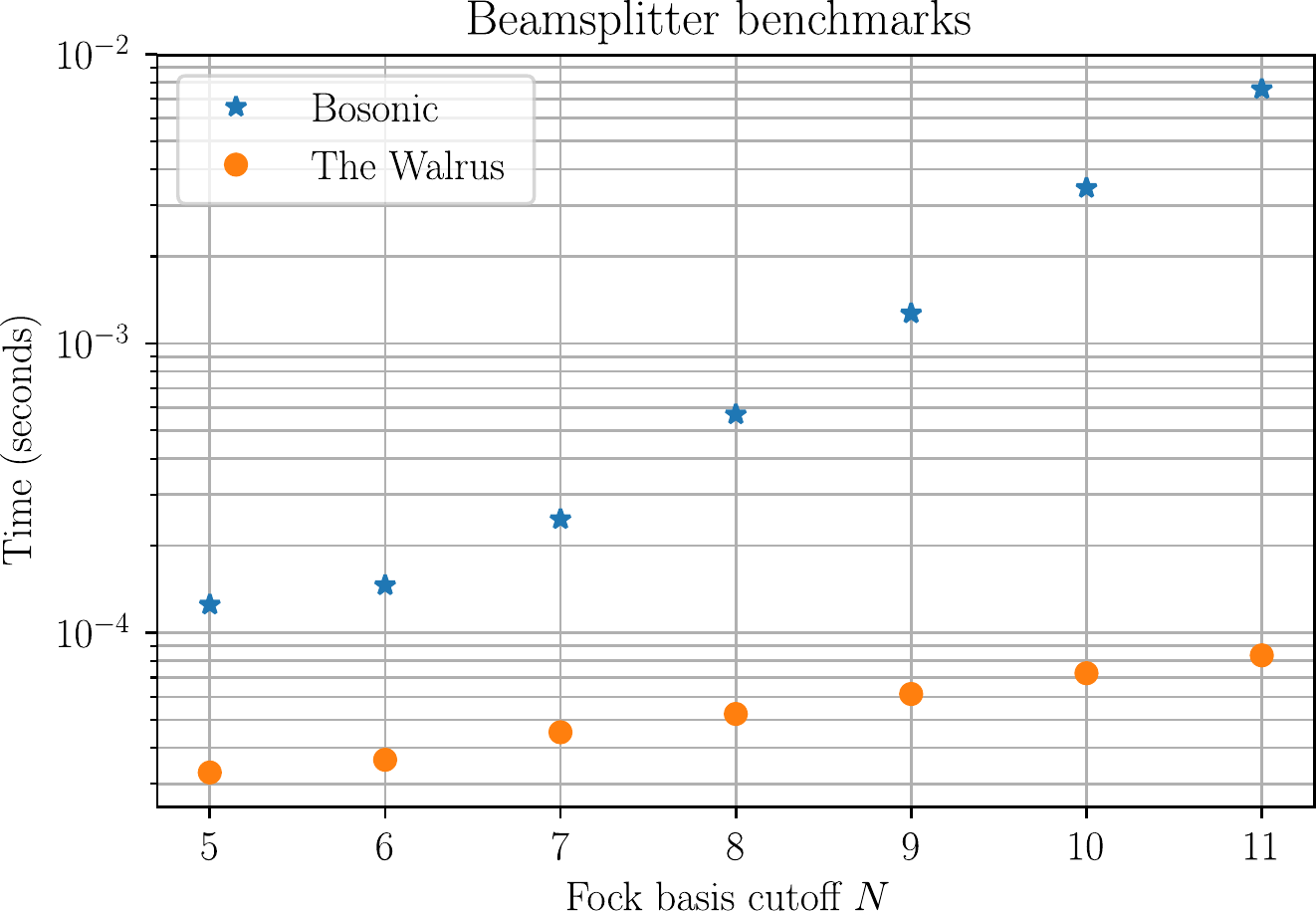}
	\caption{Benchmarks for the computation of the beamsplitter $BS(\theta, \varphi)$ using  \texttt{Bosonic}  and the methods from this work as implemented in \texttt{The Walrus}. Note that for a given cutoff $N$, \texttt{Bosonic} calculates $\mathcal{B}^{(N)}_{i,j} = \braket{N-i,i|BS(\theta,\varphi)|N-j,j}$ and that in their implementation the constraint $N \leq 11$ is hard-coded. For a fair comparison we benchmark their code against a wrapper function that calculates the tensor $\braket{m,n|BS(\theta,\varphi)|k,l}$ and then constructs the relevant matrix $\mathcal{B}^{(N)}_{i,j}$ from it. Note that our methods offer a very significant improvement in speed even for the modest cutoff considered here.
		The benchmarks in this figure were performed on a single core of an Intel-i5 processor @ 2.50 GHz.
	}
	\label{fig:bosonicvstw}
\end{figure}

Second, we compare our methods against the implementations of the beamsplitter in the package \texttt{Bosonic}\cite{steinbrecher2019quantum}. This package has a hard-coded cutoff of $11$. We benchmark up to this cutoff in Fig.~\ref{fig:bosonicvstw}.
Even for these modest sizes our implementation achieves almost two orders of magnitude in improvement.

\begin{table*}[!ht]
	\centering
	\begin{tabular}{lccc}
		\hline
		\hline
		\emph{Hyperparams.} & \emph{Single photon} & \emph{ON} & \emph{Hex GKP} \\
		\hline
		Cutoff $N$ & 100 (8) & 100 (14) & 100 (50)\\
		Layers $M$ & 8 & 20 & 35 (25)\\
		GD Steps &1500 (5000) & 2500 (5000) & 5000 (10000) \\
		\hline
		\emph{Results}\\
		\hline
		Fidelity & 99.998\% & 99.995 (99.93)\% & 99.83 (99.60)\%\\
		Runtime (s) & 50 (65) & 260 (436) & 720 (6668) \\
		\hline
		\hline
	\end{tabular}
	\caption{Optimization results for state generation. We indicate in parenthesis the values from Table I in Ref.  \cite{arrazola2019machine} if they differ from ours. Thanks to the improved speed of our recursive algorithm we can push the cutoff to $N=100$ and we can increase the number of layers for the generation of the Hex GKP state. Compared to the results in Table I of Ref.  \cite{arrazola2019machine} we can either reach the same fidelity with fewer steps or overall better results (higher fidelity, fewer steps, shorter runtime). All of these optimizations have been run on a single core of an Intel Core i5 @ 3,1 GHz. For comparison the results in Ref.~\cite{arrazola2019machine} are obtained in hardware of comparable speed for the first two target states and on a 20-core Intel Xeon CPU @ 2.4GHz with 252GB RAM for the Hex GKP state.}
	\label{tab:states}
\end{table*}

Third, we test a circuit optimization task. We start by setting up a circuit built by alternating single-mode Gaussian gates and single-mode Kerr gates. A Kerr gate is a non-linear optical interaction which varies the phase of a Fock state according to the square of the occupation number:
\begin{align}
K(\kappa) = \exp\left[i\kappa (a^\dagger a)^2 \right],
\end{align}
and is therefore diagonal in the photon number basis. 
Such single-mode circuit corresponds to a unitary transformation which we can write as a stack of $M$ layers:
\begin{align}
U(\bm \gamma, \bm \phi, \bm \zeta, \bm{\kappa}) = \prod_{m = 1}^M \mathcal{G}^{(1)}(\gamma_m, \phi_m, \zeta_m)K(\kappa_m).
\end{align}
We then input the single-mode vacuum state $|0\rangle$ and optimize the parameters such that the unitary $U(\bm \gamma, \bm \phi, \bm \zeta)$ produces an output as close as possible to the desired target state. We choose three target states: a single photon state $|1\rangle$, an ``ON'' state, i.e. a superposition of the vacuum and a Fock state, in our case $(|0\rangle + |9\rangle)/\sqrt{2}$ and finally a Hex GKP state, which is a logical state of the single-mode GKP error correcting code \cite{gottesman2001encoding}.
To evaluate the quality of the state produced by the circuit we define a loss function between the output $U|0\rangle$ and the target:
\begin{align}
L(|\psi_\mathrm{out}\rangle) = -|\langle\psi_\mathrm{out}|\psi_\mathrm{target}\rangle|.
\end{align}
For the optimization we run the adaptive gradient descent algorithm \verb|Adam|. We note that the range over which the parameters are optimized is unconstrained, except for the fact that the parameters $\phi_m$ and $\kappa_m$ are real.

The results are presented in Table \ref{tab:states} and can be compared with the ones in table I of Ref.~\cite{arrazola2019machine} (reported in parenthesis where they differ from ours). Our method allows us to use much higher cutoff dimensions, leading to equal or better fidelities in a significantly shorter time or fewer gradient descent steps (GD Steps in the table). The optimization of the Hex GKP state is particularly striking, as it reached better results in a fraction of the time while running on much slower hardware. 

We point out that even if the final output has to contain few photons, using a large cutoff is beneficial because it allows intermediate layers to explore higher excitation numbers, which may be necessary to reach the desired output with high fidelity.

\section{Conclusions}
We have presented a unified procedure for obtaining the Fock representation of a quantum optical Gaussian transformation using generating functions. From the generating functions we obtain recurrence relations which allow us to compute the matrix elements of any given Gaussian transformation when written in terms of its Bloch-Messiah decomposition. Furthermore, we have developed general techniques to also obtain the gradients of a given transformation once it is parametrized.

As particular examples we derived explicit recurrence relations for the displacement, single- and two-mode squeezing gates and the beamsplitter. 
These recurrence relations are part of the \verb|fock_gradients| module of \texttt{The Walrus} written in pure Python using \texttt{Numpy} and are sped up using the just-in-time compiling capabilities of \texttt{Numba}. Our highly portable implementation makes our algorithms ideal for high performance simulation of quantum optical circuits using both CPUs and GPUs. Furthermore, we expect them to accelerate the research on quantum hardware, quantum machine learning, optical data processing, device discovery and device design.

\emph{Note added --} While preparing this manuscript we became aware of related work by J. Huh \cite{huh2020multimode}.

\section*{Acknowledgements}
N.Q. thanks Safwan Hossain, Theodor Isacsson, Josh Izaac, Nathan Killoran, John E. Sipe and Antal Sz\'ava for valuable discussions.
F.M. thanks Electra Eleftheriadou for support throughout this work, Yuan Yao for helping with the correctness of the final version of the manuscript, Fr\'ed\'eric Grosshans, Steve Barnett, Gerardo Adesso and Alessio Serafini for helpful inputs.
The authors also thank the open source scientific computing community, in particular the developers of \texttt{Numpy} \cite{walt2011numpy}, \texttt{Scipy} \cite{virtanen2020scipy}, \texttt{Jupyter} \cite{kluyver2016jupyter},
\texttt{Matplotlib} \cite{hunter2007matplotlib} and \texttt{Numba} \cite{lam2015numba}, without whom this research would not have been possible.

\onecolumngrid
\appendix

\section{Multimode generating function}\label{app:gaussian}
In Barnett and Radmore \cite{barnett2002methods} (cf. Sec. 3.7 and Appendix 5) the following identities are proven:
\eq{
	\exp\left(\theta  a^\dagger  a\right) &= \ :\exp\left( \left[\exp(\theta)-1 \right]  a^\dagger  a \right): =\sum_{n=0}^\infty \frac{\left[\exp(\theta)-1 \right]^n \a^{\dagger n} \a^n}{n!},
}
and also
\eq{
	{S}(\zeta) =& 	\exp\left(\frac12 \zeta^*  a^2 -\hc   \right) =	\exp\left(\frac12 r e^{-i \delta}  a^2 -\hc   \right) \nn \\
	=& \exp\left(-\tfrac{\tanh r}{2} e^{i \delta}  a^{\dagger 2} \right)  \exp\left(-\left[ a^\dagger  a +\tfrac{1}{2} \right] \log \cosh r\right)   \exp\left(\tfrac{\tanh r}{2} e^{i \delta}  a^{ 2} \right).
} 
where we wrote $\zeta = r e^{ i\delta}$. 
Putting these two identities together we can normal order a squeezing operator
\eq{
	\exp\left(\frac12 r e^{-i \delta}  a^2-\hc  \right)=	& \exp\left(-\tfrac{\tanh r}{2} e^{i \delta}  a^{\dagger 2} \right)  \times \frac{: \exp\left(  \left[ \sech  r -1 \right] \  a^\dagger  a\right):}{\sqrt{\cosh r}}   \times \exp\left(\tfrac{\tanh r}{2} e^{-i \delta}  a^{ 2} \right). 
}
Using this expression one can find the matrix element of the squeezing operation between two coherent states
\eq{
	&	\bra{\alpha^*}\exp\left(\frac12 r e^{-i \delta}  a^2-\hc  \right)	\ket{\beta}=\frac{1}{\sqrt{\cosh r}} \braket{\alpha^*|\beta} \exp\left(-\tfrac{\tanh r}{2} e^{i \delta} \alpha^{2}+ \left[ \sech  r -1 \right] \ \alpha \beta+ \tfrac{\tanh r}{2} e^{-i \delta} \beta^{ 2} \right), 
}
and noting that
\eq{
	\braket{\alpha^*|\beta} = \exp\left(-\tfrac{1}{2} \left[|\alpha|^2+|\beta|^2 - 2 \alpha \beta  \right]\right),	
}
one can write
\eq{
	\bra{\alpha^*}&\exp\left(\frac12 r e^{i \delta}  a^2 -\hc  \right)	\ket{\beta}= 	\frac{ \exp\left(-\tfrac{1}{2} \left[|\alpha|^2+|\beta|^2  \right]\right)}{\sqrt{\cosh r}}  \exp\left( - \frac12\left[\alpha \  \beta \right] 
	\left[
	\begin{array}{cc}
		e^{i \delta}\tanh r & -\sech  r  \\
		-\sech  r  & -e^{-i \delta} \tanh r \\
	\end{array}
	\right]
	\left[
	\begin{array}{c}
		\alpha \\
		\beta  \\
	\end{array}
	\right]
	\right).
}
This proof generalizes in a straightforward way to
\eq{\label{eq:bigsq}
	&\bra{\bm{\alpha}^*}\mathcal{ S}(\bm{\zeta})	\ket{\bm{\beta}}= 	 \frac{\exp\left(-\tfrac{1}{2} \left[||\bm{\alpha}||^2+||\bm{\beta}||^2  \right]\right)}{\sqrt{\prod_{i=1}^\ell \cosh r_i}} \exp\left( -\frac12\left[\bm{\alpha} \  \bm{\beta} \right] 
	\left[
	\begin{array}{c|c}
		\diag(e^{i \bm{\delta}}\tanh \bm{r}) & -\diag(\sech  \bm{r} )  \\
		\hline
		-\diag(\sech  \bm{r} )  & -\diag(e^{-i \bm{\delta}}\tanh \bm{r}) \\
	\end{array}
	\right]
	\left[
	\begin{array}{c}
		\bm{\alpha} \\
		\bm{\beta}  \\
	\end{array}
	\right]
	\right),	\nn
}
where $\zeta_j = r_j e^{i \delta_j }$.

From the definitions in Eq.~\eqref{mapping} it is direct to show that
\eq{
	\mathcal{ U}(\bm{V}) \ket{\bm{\beta}} = \ket{ \bm{V} \bm{\beta}} \text{ and }	\bra{\bm{\alpha}^*}  \mathcal{ U}(\bm{W}) = \bra{  \bm{W}^\dagger \bm{\alpha}^* }.
}
Finally, let us note that
\eq{ \label{eq:disp}
	\bra{\bm{\alpha}^*} \mathcal{ D}	(\bm{\gamma}) = \bra{\bm{\alpha}^* - \bm{\gamma}} \exp\left(\tfrac{1}{2} \left[ \bm{\alpha}^T \bm{\gamma}-\bm{\gamma}^\dagger \bm{\alpha}^* \right] \right).
}
To show this last identity recall that $\bra{\bm{\alpha}^*}  = \bra{\bm{0}} \mathcal{ D}(-\bm{\alpha}^*)$  and use the composition rule for displacement operators (cf. Eq. 3.6.30 of Ref.~\cite{barnett2002methods}).
With this setup we are finally ready to prove Eq.~\ref{eq:exp} as follows:
\seq{
	\eq{
		\braket{\bm{\alpha}^* &| \mathcal{ D }(\bm{\gamma}) \mathcal{ U}(\bm{W}) \mathcal{ S }(\bm{\zeta}) \mathcal{ U}(\bm{V}) 	| \bm{\beta}} \\
		&	 = \exp\left(\tfrac{1}{2} \left[ \bm{\alpha}^T \bm{\gamma}-\bm{\gamma}^\dagger \bm{\alpha}^* \right] \right)\braket{\bm{\alpha}^* -\bm{\gamma}|  \mathcal{ U}(\bm{W}) \mathcal{ S }(\bm{\zeta}) 	| \bm{V}\bm{\beta}}  \\
		&	 = \exp\left(\tfrac{1}{2} \left[ \bm{\alpha}^T \bm{\gamma}-\bm{\gamma}^\dagger \bm{\alpha}^* \right] \right)\braket{ \bm{W}^\dagger (\bm{\alpha}^* -\bm{\gamma})|   \mathcal{ S }(\bm{\zeta}) 	| \bm{V}\bm{\beta}} \\
		&=\frac{\exp\left(\tfrac{1}{2} \left[ \bm{\alpha}^T \bm{\gamma}-\bm{\gamma}^\dagger \bm{\alpha}^* \right] \right)}{\sqrt{\prod_{i=1}^\ell \cosh r_i}} \exp\left(-\tfrac{1}{2} \left[ ||\bm{\alpha}^*-\bm{\gamma}||^2 +|\bm{\beta}|^2\right]\right) \times \\
		& \quad \quad 
		\exp\left( -\tfrac{1}{2}\left[( \bm{\alpha}^T -\bm{\gamma}^\dagger ) \bm{W}, \ \  \bm{\beta}^T \bm{V}^T \right] 
		\left[
		\begin{array}{c|c}
			\diag(e^{i \bm{\delta}} \tanh \bm{r}) & -\diag(\sech  \bm{r} )  \\
			\hline
			-\diag(\sech  \bm{r} )  & -\diag(e^{-i \bm{\delta}} \tanh \bm{r}) \\
		\end{array}
		\right]
		\left[
		\begin{array}{c}
			\bm{W}^T(\bm{\alpha}-\bm{\gamma}^*) \\
			\bm{V}	\bm{\beta}  \\
		\end{array}
		\right]
		\right). \nn
	}
}
At this point we introduce the symmetric, complex-unitary matrix
\eq{
	\bm{\Sigma} &= 	\left[
	\begin{array}{c|c}
		\bm{W} \diag(e^{i \bm{\delta}} \tanh \bm{r}) \bm{W}^T &  	-\bm{W} \diag(\sech  \bm{r}) \bm{V}\\
		\hline
		-\bm{V}^T \diag(\sech  \bm{r}) \bm{W}^T & - \bm{V}^T \diag(e^{-i \bm{\delta}}\tanh \bm{r}) \bm{V}
	\end{array}
	\right] \\
	&= 
	\left[
	\begin{array}{c|c}
		\bm{W}  &  	\bm{0} \\
		\hline
		\bm{0}  & \bm{V}^T 
	\end{array}
	\right]	
	\left[
	\begin{array}{c|c}
		\diag(e^{i \bm{\delta}} \tanh \bm{r})  &  	 -\diag(\sech  \bm{r})  \\
		\hline
		-\diag(\sech  \bm{r}) & -\diag(e^{-i \bm{\delta}} \tanh \bm{r})
	\end{array}
	\right]	\left[
	\begin{array}{c|c}
		\bm{W}  &  	\bm{0} \\
		\hline
		\bm{0}  & \bm{V}^T 
	\end{array}
	\right]	^T,
}
and the complex vector
\eq{
	\bm{\nu} &= \left[ \begin{array}{c}
		\bm{\alpha} \\
		\bm{\beta}
	\end{array} \right] \in \mathbb{C}^{2 \ell},
}
to finally write
\eq{
	\braket{\bm{\alpha}^* | \mathcal{ D }(\bm{\gamma}) \mathcal{ U}(\bm{W}) \mathcal{ S }(\bm{\zeta}) \mathcal{ U}(\bm{V})| \bm{\beta}} = 	
	\exp\left(-\tfrac{1}{2} \left[ ||\bm{\alpha}||^2 + ||\bm{\beta}||^2  \right] \right)C \exp\left( \bm{\mu}^T \bm{\nu} - \frac12 \bm{\nu}^T \bm{\Sigma} \bm{\nu} \right),
}
where
\eq{
	C &= \frac{ \exp\left(-\tfrac{1}{2} \left[  ||\bm{\gamma}||^2  + \bm{\gamma}^\dagger  \bm{W} \diag(e^{i \bm{\delta}}  \tanh \bm{r}) \bm{W}^T  \bm{\gamma}^*\right]\right)}{\sqrt{\prod_{i=1}^\ell \cosh r_i}} , \\
	\bm{\mu}^T &= \left[\bm{\gamma}^\dagger \bm{W} \diag(e^{i \bm{\delta}}  \tanh \bm{r}) \bm{W}^T+\bm{\gamma}^T, -\bm{\gamma}^\dagger \bm{W} \diag(\sech  \bm{r}) \bm{V} \right]	,
}
which is the expression quoted in the main text.
Note that a related expression for the matrix elements of a general Gaussian transformations evaluated in the coordinate basis was worked out by Moshinsky and Quesne \cite{moshinsky1971linear}.
\section{Recurrence relation for the high-order derivatives of an exponential function}\label{app:rec}

We rewrite Eq.~(15) of Ref.~\cite{miatto2019recursive} in our notation as follows:
\eq{\label{main}
	\Gamma_{\bm{k}+1_n} = \sum_{\bm{j}=\bm{0}}^{\bm{k}}	\binom{\bm{k}}{\bm{j}} Q^{(\bm{k}-\bm{j} +1_n)} \Gamma_{\bm{j}},
}
where we write $\partial_{\bm \nu}^{\bm{k}} f(\bm{\nu})|_{\bm{\nu}=\bm{0}}  = f^{(\bm{k})} = f_{\bm{k}}  $ for the $\bm{k}^{\text{th}}$ derivative of $f$ at zero and define $\sum_{\bm{j}=\bm{0}}^{\bm{k}}	\binom{\bm{k}}{\bm{j}} = \sum_{j_1}^{k_1}\dots\sum_{j_n}^{k_n}\binom{k_1}{j_1}\dots \binom{k_n}{j_n}$.
In our case $Q$ is a quadratic in $\bm{\nu}$ and we write
\eq{\label{quad}
	Q(\bm{\nu}) = \bm{\mu}^T \bm{\nu}-\frac{1}{2} \bm{\nu}^T \bm{\Sigma} \bm{\nu}.
}
We can write easily all its nonzero derivatives:
\eq{
	Q^{(1_i)} &= \mu_i,\\
	Q^{(1_i+1_j)} &= -\Sigma_{i,j}.
}
We know that most of the terms in Eq.~\eqref{main} will not survive the summation, indeed only the ones that satisfy $\bm{k}-\bm{j} = 1_i$ or $\bm{k}-\bm{j} = 1_i+1_l$ will contribute to the sum. We can use this to our advantage and write 
\eq{\label{rec}
	\Gamma_{\bm{k}+1_n} &= 	 Q^{(1_n)}\Gamma_{\bm{k}} + 
	\sum_{l} k_l Q^{(1_n+1_l)} \Gamma_{\bm{k}-1_l}
	= 		\mu_n  \Gamma_{\bm{k}} - 
	\sum_{l} k_l \Sigma_{l,n} \Gamma_{\bm{k}-1_l}.
}

Now let us derive this in a different way. Start with the generating function of the multidimensional Hermite polynomials \cite{berkowitz1970calculation, kok2001multi, mizrahi1975generalized}:
\eq{\label{eq:hermitegen}
	\exp\left( \bm{\mu}^T \bm{\nu} - \tfrac{1}{2}\bm{\nu}^T \bm{\Sigma} \bm{\nu}\right)
	= \sum_{\bm{k} = \bm{0}}^{\bm{\infty}} \frac{\bm{\nu}^{\bm{k}}}{\bm{k}!} G_{\bm{k}}^{(\bm{\Sigma})}(\bm{\mu}).
}
They satisfy the recurrence relation
\eq{\label{eq:recur}
	G_{\bm{k}+1_i}^{(\bm{\bm{\Sigma}})}(\bm{\mu}) =  \mu_i  G_{\bm{k}}^{(\bm{\bm{\Sigma}})}(\bm{\mu}) 	- \sum_{j=1} \Sigma_{i,j} k_j G_{\bm{k}-1_j}^{(\bm{\bm{\Sigma}})}(\bm{\mu}), \nn
}
from which we identify that for a quadratic function $Q$,
\eq{
	\Gamma_{\bm{k}} = G^{(\bm{\Sigma})}_{\bm{k}}(\bm{\mu}) = \sqrt{ \bm{k}!} \mathcal{G}_{\bm{k} 	}.
}
By replacing this last expression in the recurrence relation Eq.~\eqref{rec}, and simplifying the factorials one arrives at the recurrence relation Eq.~\eqref{eq:recrel} for the matrix elements $\mathcal{G}_{\bm{k}}$.


\section{Recursive matrix elements from commutation relations
}\label{sec:comms}
In this section we use an alternative approach for obtaining the matrix elements and gradients of a gate by using the commutation relations with the destruction operator of a given mode. To illustrate this method we will focus on the $k^{\text{th}}$-order-phase gate defined as
\eq{\label{eq:kphase}
	V^{(k)}(\eta) = \exp\left( i \frac{\eta}{ k \hbar}  q^k \right), 
}
where the canonical position is defined as $q=\sqrt{\frac{\hbar}{2}} (\a+\ad)$ and its canonical momentum is given by $p  = i \sqrt{\tfrac{\hbar}{2}} \left( \ad-\a \right)$. In the Heisenberg picture the  $k^{\text{th}}$-order-phase gate transforms the operators according to
\eq{
	V^{(k)\dagger}(\eta) q  V^{(k)}(\eta) &= q, \\
	V^{(k)\dagger}(\eta) p  V^{(k)}(\eta) &= p + \eta q^{k-1},\\
	V^{(k)\dagger}(\eta) a  V^{(k)}(\eta) &= a+ i  \frac{\eta}{\hbar} \left( \frac{\hbar}{2}\right)^{k/2} \left(a+a^\dagger\right)^{k-1}. \label{eq:commV}
}
One can write the matrix elements of the $k^{\text{th}}$-order-phase gate in terms of integrals as follows
\eq{\label{eq:integral}
	\braket{m|&V^{(k)}(\eta)|n} = V_{m,n}^{(k)} =  \frac{1}{\sqrt{\pi 2^{n+m} n! m!}} 
	\int dx \exp\left(-x^2+ i \frac{\eta \hbar^{k/2}}{\hbar k } x^k  \right) ~ H_n(x) H_m(x),
}
where $H_{n}(x)$ is a Hermite polynomial of degree $n$ \cite{barnett2002methods}. The last equation makes explicit that the matrix elements $V_{m,n}^{(k)}$  are symmetric $V_{m,n}^{(k)} = V_{n,m}^{(k)}$ and moreover, given the parity of the Hermite polynomials $H_{n}(-x) = (-1)^{n}H_n(x)$, that the $k^{\text{th}}$-order-phase gate also has parity, i.e.,
\eq{
	V_{m,n}^{(k)}(\eta) &= (-1)^{m+n} V_{m,n}^{(k)}(-\eta) \text{ if $k$ is odd,}\\
	V_{m,n}^{(k)}(\eta) &= 0 \text{ if $k$ is even and $n+m$ is odd.}
}
Finally, note that the matrix elements of the conjugate $k^{\text{th}}$-order-phase gate \eq{
	\tilde{V}^{(k)}(\eta) = \exp\left( i \frac{\eta}{  \hbar k}  p^k \right) ,
}
are related to the ones for the $k^{\text{th}}$-order-phase gate via 
\eq{
	\braket{m|\tilde{V}^{(k)}(\eta)|n} = \tilde{V}^{(k)}_{m,n} = (i)^{m+n} V_{m,n}^{(k)},
}
which can be easily confirmed by noticing that the rotation gate by angle $\theta = \pi/2$ transforms $q$ into $p$ in the Heisenberg picture and that when acted on a Fock state it gives $R(\pi/2)\ket{n} = (i)^n \ket{n}$. 

Having established these basic facts about the  $k^{\text{th}}$-order-phase gate we are now ready to write a recurrence relation for its matrix elements. To this end, we write the  following commutator
\eq{\label{eq:kcomm}
	[\a, V^{(k)}(\eta) ] = \a V^{(k)}(\eta) - V^{(k)}(\eta) \a = V^{(k)}(\eta)V^{(k) \dagger}(\eta)  \a V^{(k)}(\eta) - V^{(k)}(\eta) \a = V^{(k)} \left\{ i  \frac{\eta}{\hbar} \left( \frac{\hbar}{2}\right)^{k/2} (\a+\ad)^{k-1} \right\}  .
}
If we multiply the last equation on the left by $\bra{m}$ and on the right $\ket{n}$ we obtain
\eq{\label{eq:kphase2}
	\sqrt{m+1} V^{(k)}_{m+1,n} - \sqrt{n} V^{(k)}_{m,n-1} = \bra{m}V^{(k)}(\eta) \left(  i  \frac{\eta}{\hbar} \left( \frac{\hbar}{2}\right)^{k/2} (\a+\ad)^{k-1} \right) \ket{n}. 
}
Note that for any $k$ one can expand the right hand side of the last equation and write 
$ (\a+\ad)^{k-1} \ket{n} = \sum_{k=n-(k-1)}^{n+k-1} c_{k} \ket{k+n}$ allowing us to obtain the desired recurrence relation. In the next subsection we specialize these results for the values $k=3$ and $k=4$. Note that the cases $k=1,2$ are Gaussian gates and thus can be dealt with using the methods of the main text.

\subsection{The cubic-phase gate}
The cubic phase gate, introduced in Ref. \cite{gottesman2001encoding}, is the simplest non-Gaussian operation one can consider when studying polynomial generators of continuous-variable quantum gates. This gate is obtained by setting $k=3$ in Eq.~\eqref{eq:kphase} and can in principle be used to decompose any higher order gate generated by a polynomial in the canonical position $q$ and momentum $p$ of a continuous-variable system by employing approximate commutator expansions \cite{lloyd1999quantum, braunstein2005quantum, weedbrook2012gaussian} or, for many important applications, finding exact decompositions \cite{kalajdzievski2018continuous,kalajdzievski2019exact}. 
Significant effort has been directed into realizing this gate either by using optical nonlinearities \cite{yanagimoto2019engineering} or measurements \cite{gottesman2001encoding,gu2009quantum,marshall2015repeat,sabapathy2018states,sabapathy2019production}. 

To obtain the recurrence relation for the cubic phase gate, we set $k=3$ in Eq.~\eqref{eq:kphase2} and rearrange to obtain
\eq{\label{eq:reccubic}
	V^{(3)}_{m,n+2} = \frac{1}{\sqrt{(n+1)(n+2)}} \biggl( \frac{i 2 \sqrt{2}}{\sqrt{\hbar} \eta} \left\{ \sqrt{n} V^{(3)}_{m,n-1} -\sqrt{m+1} V^{(3)}_{m+1,n} \right\}  - (2n+1) V^{(3)}_{m,n} - \sqrt{n(n-1)} V^{(3)}_{m,n-2} 
	\biggr).
}
We can obtain the first two rows of $V_{m,n}^{(3)}$ by setting $m=0$ or $1$ in the last equation and using the following initial seeds (cf. Appendix \ref{app:cubic})
\eq{
	V_{0,0}^{(3)} &= 2 \sqrt{\pi } e^{\frac{2 y ^6}{3}} |y|
	\text{Ai}\left(y ^4\right),  \\
	V_{1,1}^{(3)} &= -2 i \sqrt{2} y^3 V_{0,1}^{(3)} = -2 i \sqrt{2} y ^3 V_{1,0}^{(3)}=-8 \sqrt{\pi } e^{\frac{2 y^6}{3}} |y|^5
	\left(\text{Ai}'\left(y^4\right)+y^2
	\text{Ai}\left(y^4\right)\right),
}
where $y^3 = 1 / (\sqrt{\hbar} \eta )$ and $\text{Ai}(x)$ and $\text{Ai}'(x)$ are the Airy function and its first derivative \cite{abramowitz1948handbook,olivier2010airy}.
Since $V^{(k)}_{m,n}$ is symmetric in $m,n$ once the first two rows are obtained one also obtains the first two columns which serve as the initial conditions to recursively obtain any other row and thus the whole matrix. 

Even though the numerical recursive relation derived are exact, they are not numerically stable in the regime where $\eta<1$; notice that any small numerical error in the difference $\sqrt{n} V^{(3)}_{m,n-1} -\sqrt{m+1} V^{(3)}_{m+1,n} $ inside the curly braces in Eq.~\ref{eq:reccubic} will be \emph{amplified} by a factor of $1/\eta$ that will cause an uncotrollable loss of precision; because of this, the recurrence relation can only be safely used in the regime where $\eta>1$.

\subsection{The quartic-phase gate}
The quartic phase interaction arises in the context of self-interactions in scalar quantum field theories \cite{marshall2015quantum, zee2010quantum} and can also be use to achieve universality in GKP encoded qubits\cite{garcia2020efficient,gottesman2001encoding}. We can obtain a recursive relation for the matrix elements of this gate by setting $k=4$ in Eq.~\eqref{eq:kphase2} to obtain
\eq{\label{eq:recquartic}
	V^{(4)}_{m+3,n} = \frac{1}{\sqrt{(n+1)(n+2)(n+3)}}\biggl( &
	\frac{4 i }{ \eta \hbar} \left\{ \sqrt{n} V^{(4)}_{m,n-1} - \sqrt{m+1} V^{(4)}_{m+1,n}  \right\} -\sqrt{n(n-1)(n-2)} V^{(4)}_{m,n-3} \nn \\
	&- 3 n^{\tfrac{3}{2}} V^{(4)}_{m,n-1} - 3 (n+1)^{\tfrac32} V^{(4)}_{m,n+1}
	\biggr).
}
We can iterate this relation using the following initial values
\eq{
	V^{(4)}_{0,0} =& \left.  f(w,\lambda) \right|_{\lambda=1},  \\ 
	V^{(4)}_{1,1} =& \left. -2 \frac{\partial}{\partial \lambda} f(w,\lambda) \right|_{\lambda=1},\\ 
	V_{0,2}^{(4)} =&V_{2,0}^{(4)} =  \frac{1}{\sqrt{2}} \left( V_{1,1}^{(4)} - V_{0,0}^{(4)} \right), \\
	V_{2,2}^{(4)} =& \frac{1}{2}V_{0,0}^{(4)}- V_{1,1}^{(4)} +  2 \left. \frac{\partial^2}{\partial \lambda^2} f(w,\lambda) \right|_{\lambda=1},
}
where $f(w,\lambda)$ is defined in Eq.~\eqref{eq:fdef} and recalling that $V_{0,1}^{(4)} =  V_{1,0}^{(4)} = V_{1,2}^{(4)}=V_{2,1}^{(4)}=0$ by parity.  With these nine initial values we can fill the first three rows of $V_{m,n}^{(4)}$ using the recurrence relation. Then we can use the symmetry to fill the first three columns and finally complete the whole matrix. 
Note however that, like its cubic counterpart, the recurrence relation in Eq.~\eqref{eq:recquartic} is numerically unstable for $\eta<1$.

\section{Analytic matrix elements of the cubic-phase gate}\label{app:cubic}

In this appendix we obtain the matrix elements of the cubic-phase gate. We start from Eq.~\eqref{eq:integral} with  $k=3$ and $y^3 = 1/(\sqrt{\hbar} \eta)$ and doing the change of variables $x \to y k - i y^3$, and assuming without loss of generality that $y>0$,  we obtain
\eq{
	V^{(3)}_{m,n} = \frac{y e^{\frac{2}{3} y^6}}{\sqrt{\pi 2^{n+m} n! m!}} \int_{-\infty}^{\infty} dk \exp\left( i y^4 k +i k^3/3 \right)~ H_n( y k - i  y^3 ) H_m( y k - i  y^3).
}
To make progress we recall the identity $	H_n(x+y) = \sum_{k=0}^n \binom{n}{k}H_{k}(x) (2y)^{(n-k)}$, to obtain
\eq{
	V^{(3)}_{m,n} &= 		\frac{y e^{\frac{2}{3} y^6}}{\sqrt{\pi 2^{n+m} n! m!}} 
	\int_{-\infty}^{\infty} dk \exp\left( i y^4 k +i k^3/3 \right)~ 
	\sum_{k=0}^n \binom{n}{k}H_{k}(-i y^3) (2 y k)^{(n-k)}
	\sum_{l=0}^m \binom{m}{l}H_{l}(-i y^3) (2 y k)^{(m-l)} \\
	&= 		\frac{y e^{\frac{2}{3} y^6}}{\sqrt{\pi 2^{n+m} n! m!}} \sum_{k=0}^n  \sum_{l=0}^m (2 y)^{m+n-k-l} \binom{n}{k}H_{k}(-i y^3)  \binom{m}{l}H_{l}(-i y^3) \int_{-\infty}^{\infty} dk \exp\left( i y^4 k +i k^3/3 \right)   k^{n+m-k-l}\\ 
	&= 		\frac{2 \sqrt{\pi} y e^{\frac{2}{3} y^6}}{\sqrt{ 2^{n+m} n! m!}} \sum_{k=0}^n  \sum_{l=0}^m (-2 i y)^{m+n-k-l} \binom{n}{k}H_{k}(-i y^3)  \binom{m}{l}H_{l}(-i y^3) 
	\Bigg. \left[ \frac{d^{n+m-l-k}}{dw^{n+m-l-k}} \Ai(w) \right]\Bigg|_{w=y^4}. 	
}
In the last equation, we used the well known Fourier expansion of the Airy function to write~\cite{olivier2010airy}
\eq{
	\Ai(x) &= \frac{1}{2 \pi}\int_{-\infty}^{\infty} dk  \exp\left( i \frac{k^3}{3} + i k x \right),\\
	\frac{d^\ell}{dx^\ell} \Ai(x) &= \frac{1}{2 \pi}\int_{-\infty}^{\infty} dk  (i  k)^\ell \exp\left( i \frac{k^3}{3} + i k x \right).
}
Note that any higher order derivative of the Airy function can be expressed in terms of polynomials of its arguments, the Airy function itself and it first derivative as follows
\eq{
	\frac{d^\ell}{dx^\ell} \Ai(x) = p_{\ell}(x) \Ai(x)+ q_{\ell}(x)\Aip(x)	,
}
where the polynomials $p_\ell$ and $q_\ell$ satisfy the recurrence relations
\eq{\label{eq:recairy}
	p_{\ell+1}(x) = x q_\ell(x)	 +\frac{d}{dx} p_\ell(x), \quad q_{\ell+1}(x) = p_\ell(x)+\frac{d}{dx}q_\ell(x).	
}
By iterating the recurrence relations in Eq.~\eqref{eq:recairy} one can in principle obtain the polynomials $p_{\ell}(x)$ and $q_\ell(x)$ and then any derivative of the Airy function and thus any matrix element $V^{(3)}_{n,m}$.

\section{Analytic matrix elements of the quartic-phase gate}\label{app:quartic}

We are now interested in calculating
\eq{
	V^{(4)}_{m,n}=\braket{m|V^{(4)}(\eta)|n} = \frac{1}{\sqrt{\pi 2^{n+m} n! m!}}  \int_{-\infty}^{\infty} dx    \exp\left(-x^2 +    \tfrac{i}{4} w  x^4 \right)~ H_n(x) H_m(x), \nn 	 
}
where $H_n(x)$ is the $n^{\text{th}}$ order Hermite polynomial and we introduced $w = \hbar \eta $.
First consider the related integral (cf. Eq.~3.968 of Ref.~\cite{gradshteyn1980table})
\eq{\label{eq:fdef}
	f(w, \lambda) &= \frac{1}{\sqrt{\pi}} \int_{-\infty}^{\infty} dx    \exp\left(-\lambda x^2 +    \tfrac{i}{4} w  x^4 \right) \\
	&= -\frac{i}{4} \sqrt{\pi } e^{\frac{i \lambda ^2}{8 w }+\frac{i \pi }{8}}
	\sqrt{\frac{\lambda }{w }} \left( J_{\frac{1}{4}}\left(\frac{\lambda ^2}{8 w }\right) -i Y_{\frac{1}{4}}\left(\frac{\lambda ^2}{8
		w }\right) \right) = -\frac{i}{4} \sqrt{\pi } e^{\frac{i \lambda ^2}{8 w }+\frac{i \pi }{8}}
	\sqrt{\frac{\lambda }{w }}  H^{(2)}_{\frac{1}{4}}\left(\frac{\lambda ^2}{8 w }\right),
}
where $J_{\alpha}(x)$ is a Bessel function of the first kind, $Y_{\alpha}(x)$ is a Bessel function of the second kind (sometimes also called a Neumann function) and $H_\alpha^{(2)}(x)$ is a Hankel function of the second kind (not to be confused with the Hermite polynomials $H_n(x)$). 
Note that $f(w,1) = V_{0,0}^{(4)}(w)$. To obtain any other matrix element first notice that these elements are nonzero if and only if $n+m$ is even, which implies that $H_n(x) H_m(x)$ contains only even powers of $x$. Thus we only need to know how to evaluate 
\eq{
	\frac{(-1)^k}{\sqrt{\pi }}  \int_{-\infty}^{\infty} dx \ x^{2 k}     \exp\left(-\lambda x^2 +    \tfrac{i}{4} w  x^4 \right) = \frac{\partial^k }{\partial \lambda ^k } f(w, \lambda).
}
In turn the derivatives of this function can be written as derivatives of the Hankel functions of the second kind which satisfy the recurrence relation
\eq{
	\frac{\partial }{\partial x} H_i^{(2)}(x) = \frac{1}{2} \left( H_{i-1}^{(2)}(x)+ H_{i+1}^{(2)}(x) \right).
}
Thus, any matrix element $V^{(4)}_{m,n}$ can be obtained in terms of exponentials, polynomials and Hankel functions of the second kind in the variable $w$.

\section{Gradients of single parameter gates}\label{app:single}
We outline here an alternative procedure to obtain gradients of single-parameter gates. Let us consider a gate parametrized by a single complex number $u = x e^{i \epsilon }$, 
\eq{
	G(u) = G(x e^{i \epsilon}) =  \exp\left(x e^{i\epsilon} h - \hc \right).
}
The case of a gate parametrized by a single \emph{real} number can be obtained by letting $\epsilon=\pi/2$ and writing $G(x) = \exp(i x (h+h^\dagger)) = \exp(H x)$ with $H = i (h+h^\dagger)=-H^\dagger$.
For example, for a two-mode squeezing gate, $h = a_1^\dagger a^\dagger_2$ and for the cubic phase gate, $H = i q^3/(k \hbar)$. Now it is direct to verify that
\eq{\label{eq:mod}
	\partial_x \braket{\bm{m}|G|
		\bm{n}} = \braket{\bm{m}|G  \left\{  e^{i \epsilon }h -\hc \right\} |\bm{n}}.
}
To obtain the gradient with respect to the amplitude $x$, one only needs to express $(e^{i \epsilon}h - \text{H.c.}) \ket{\bm{n}}$ in terms of a linear combination of Fock states to reduce the gradient to a linear combination of the matrix elements of the gate itself. 
Again using the two-mode squeezing gate and the cubic-phase gate as examples we easily find
\eq{
	&\partial_r \braket{\bm{m}|S^{(2)}|
		\bm{n}} = \braket{m_1, m_2|S^{(2)} \  \left\{ a_1^\dagger a_2^\dagger e^{i \epsilon } -\hc \right\}|n_1, n_2} \\
	&\quad = \sqrt{(n_1+1)(n_2+1)} e^{i \epsilon} S^{(2)}_{m_1,m_2,n_1+1,n_2+1}- \sqrt{n_1 n_2} e^{-i \epsilon} S^{(2)}_{m_1,m_2,n_1-1,n_2-1},\nn \\
	&\partial_\eta \braket{m|V^{(3)}|n} =  \braket{m| i V^{(3)}  \frac{\sqrt{\hbar}}{3 \times 2^{3/2}} (\a+\ad)^3|n}\\
	&\quad =  i \frac{\sqrt{\hbar}}{3 \times 2^{3/2}} \left( \sqrt{n(n-1)(n-2)}V^{(3)}_{m,n-3}+3 n^{\tfrac32} V^{(3)}_{m,n-1} + 3 (n+1)^{\tfrac32} V^{(3)}_{m,n+1} + \sqrt{(n+1)(n+2)(n+3)} V^{(3)}_{m,n+3} \right).\nn
}
\begin{table}[t!]
	\begin{center}
		
		\label{tab:table1}
		\begin{tabular}{c c c c}\hline
			\hline
			Gate & Parameter $u = x e^{i \epsilon}$ & Generator $h$ & Value of $s$ \\
			
			\hline
			$D(\gamma)$ & $\gamma$ & $\ad_1$ & 1\\
			$S(\zeta)$ & $\zeta=r e^{i \delta}$ & $-\tfrac{1}{2} a_1^{\dagger 2}$ & $\tfrac{1}{2}$ \\
			$S^{(2)}(\zeta)$ & $\zeta=r e^{i \delta}$ & $a_1^\dagger a_2^\dagger$ & 1 \\
			$B(\theta, \varphi)$ & $\theta e^{i \varphi}$ & $a_1 \ad_2$ & $-1$\\ \hline       \hline
		\end{tabular}
		\caption{\label{lookup}Parametrizations of different common Gaussian gates and their $s$-parameter necessary to obtain the gradient with respect to their phase.}
	\end{center}
\end{table} 

Let us consider the gradient with respect to the phase: one would hope that a formula similar to Eq.~\eqref{eq:mod} should hold. Unfortunately this is not the case, the reason being that two gates with the same phase parameter commute with each other,  $[G(x e^{i \epsilon}), G(x' e^{i \epsilon})]=0$ but the same commutativity does not hold for different values of the phase, $[G(x e^{i \epsilon}), G(x e^{i \epsilon'})]\neq 0$. To make progress we note that for any of the single complex-parameter gates one can obtain the following simple decomposition:
\eq{
	G(x e^{i \epsilon}) = R_1 (s \epsilon) G(x) R_1^\dagger(s \epsilon),
}
where $R_1(s \epsilon)=e^{i s \epsilon a_1^\dagger a_1}$ is a rotation gate. We summarize the values of the different parameters in this decomposition in Table \ref{lookup} for the single- and two-mode gates studied in the main text.
With this decomposition we can now take derivatives of matrix elements with respect to the phase argument of the complex parameter,
\eq{
	\partial_{\epsilon} \braket{\bm{m}|G|
		\bm{n}} =& \bra{\bm{m}}(\partial_{\epsilon} R_1(s\epsilon)) G(x) R_1^\dagger(\epsilon)  +  R_1 (s\epsilon) G(x) (\partial_{\epsilon} R_1^\dagger(s\epsilon))   \ket{\bm{n}} \nn \\
	=&  \bra{\bm{m}} ( i s a_1^\dagger a_1) R_1(s\epsilon) G(x) R_1^\dagger(s\epsilon) +  R_1 (s\epsilon) G(x)  R_1^\dagger(s\epsilon) (-i s a_1^\dagger a_1 )   \ket{\bm{n}} \nn  \\
	=&  i s(m_1-n_1) \braket{\bm{m}|G|\bm{n}}.
}
\twocolumn
\bibliographystyle{unsrtnat}
\bibliography{Fast_optimization_of_optical_gaussian_circuits}
\onecolumngrid

\end{document}